\documentclass[twocolumn,tighten,iop,times,twocolappendix]{aastex631}
\usepackage{amsmath}
\usepackage{iondefs, farhan-defs}
\usepackage[super]{nth}
\usepackage{bigints}
\usepackage[fulladjust]{marginnote}

\usepackage{appendix}

\hypersetup{colorlinks=true, citecolor=blue, filecolor=magenta,urlcolor=blue, linkcolor=blue}

\begin{document}


\title{\sc Evolution of {\CIV} Absorbers. II.~Where does {\CIV} live?}


\author[0000-0002-0072-0281]{Farhanul Hasan}
\affiliation{Department of Astronomy, New Mexico State University, Las Cruces, NM 88003, USA}

\author[0000-0002-9125-8159]{Christopher W. Churchill}
\affiliation{Department of Astronomy, New Mexico State University, Las Cruces, NM 88003, USA}

\author[0000-0002-6434-4684]{Bryson Stemock}
\affiliation{Department of Astronomy, New Mexico State University, Las Cruces, NM 88003, USA}

\author[0000-0003-2377-8352]{Nikole M. Nielsen}
\affiliation{Centre for Astrophysics and Supercomputing, Swinburne University of Technology, Hawthorn, Victoria 3122, Australia}
\affiliation{ARC Centre of Excellence for All Sky Astrophysics in 3 Dimensions (ASTRO 3D), Australia}

\author[0000-0003-1362-9302]{Glenn G. Kacprzak}
\affiliation{Centre for Astrophysics and Supercomputing, Swinburne University of Technology, Hawthorn, Victoria 3122, Australia}
\affiliation{ARC Centre of Excellence for All Sky Astrophysics in 3 Dimensions (ASTRO 3D), Australia}

\author[0000-0002-2908-9702]{Mark Croom}
\affiliation{Department of Astronomy, New Mexico State University, Las Cruces, NM 88003, USA}

\author[0000-0002-7040-5489]{Michael T. Murphy}
\affiliation{Centre for Astrophysics and Supercomputing, Swinburne University of Technology, Hawthorn, Victoria 3122, Australia}



\shorttitle{Where does {\CIV} live?} \shortauthors{Hasan et al.}

\begin{abstract}

We use the observed cumulative statistics of {\CIV} absorbers and dark matter halos to infer the distribution of {\CIV}-absorbing gas relative to galaxies at redshifts $0\!\leq\!z\!\leq\!5$. We compare the cosmic incidence {\dndx} of {\CIV} absorber populations and galaxy halos, finding that massive $L\!\geq\!{\Lstar}$ halos alone cannot account for all the observed {\wvweak} absorbers.  However, the {\dndx} of lower mass halos exceeds that of {\wvweak} absorbers. We also estimate the characteristic gas radius of absorbing structures required for the observed {\CIV} {\dndx}, assuming each absorber is associated with a single galaxy halo. The {\wweak} and {\wstrong} {\CIV} gas radii are $\sim$30--70\% ($\sim$20--40\%) of the virial radius of {\Lstar} ({\tenpLstar}) galaxies, and the {\wvweak} gas radius is $\sim$100--150\% ($\sim$60--100\%) of the virial radius of {\Lstar} ({\tenpLstar}) galaxies. For stronger absorbers, the gas radius relative to virial radius rises across Cosmic Noon and falls afterwards, while for weaker absorbers, the relative gas radius declines across Cosmic Noon and then dramatically rises at $z\!<\!1$. A strong luminosity-dependence of gas radius implies highly extended {\CIV} envelopes around massive galaxies before Cosmic Noon, while a luminosity-independent gas radius implies highly extended envelopes around dwarf galaxies after Cosmic Noon. From available absorber-galaxy and {\CIV} evolution data, we favor a scenario in which low-mass galaxies enrich the volume around massive galaxies at early epochs and propose that the outer halo gas ($>\!0.5{\Rv}$) was produced primarily in ancient satellite dwarf galaxy outflows, while the inner halo gas ($<\!0.5{\Rv}$) originated from the central galaxy and persists as recycled accreting gas.

\end{abstract}


\submitjournal{Astrophysical Journal}

\received{Aug 10 2021}
\revised{Oct 13 2021}
\accepted{Oct 16 2021}

\section{Introduction}

According to the standard $\Lambda$CDM paradigm of cosmology, dark matter (DM) comprises $\sim$25\% of the total energy density and $\sim$85\% of the matter density of the universe \citep{Planck20}. In overdense regions, DM gravitationally accumulates into halos and groupings and clusters of halos \citep[e.g.,][]{NFW96, Navarro10, Springel05, Guo10, Klypin11, WT18}. As baryonic gas accretes into the DM halos, the gas cools and condenses, transforming into the luminous components of the universe, including stars, galaxies, and the groups and clusters in which they amass \citep[e.g.,][]{VO04, Moster13, SD15, Naab17, Vb20}. 

Galaxy formation and evolution is largely dictated by how baryons gravitationally and hydrodynamically respond as they are cycled through DM halos. This ``baryon cycle'' not only produces the luminous components of the universe, but also generates extended gaseous structures. As much as 50\% of the baryonic mass bound to a galaxy can be contained in a highly extended surrounding region commonly referred to as the circumgalactic medium  \citep[CGM; e.g.,][]{Steidel10, Tumlinson11, Tumlinson17, Peeples14}. Additionally, large scale gaseous filaments connecting galaxies form a baryonic intergalactic medium \citep[IGM; e.g.,][]{Dave01, Mcquinn16, Danforth16, Burchett20}. 

Through the star formation and stellar feedback stages in the baryon cycle, the otherwise pristine gaseous universe is enriched by metals originally fused in stellar cores.  Starting at early cosmic epochs, galactic scale winds can eject metals, distributing them into the CGM and the IGM \cite[e.g.,][]{OD06, OD08, Oppenheimer12, Ford14, Muratov15, Muratov17, Christensen16, AA17, Hafen19, Hafen20, Nelson19}. As such, these metal enriched gaseous structures serve as diverse tracers of the baryon cycle across much of cosmic time, and, for that reason, have been widely studied in order to chart their cosmic incidence, redshift clustering, relationships to galaxy properties, kinematics, chemical and ionization conditions, and the evolution in these conditions \citep[e.g.,][]{Sargent88, Steidel90, cwc99, Nestor05, Simcoe06, Schaye07, Cooksey13, nikki13a, Bordoloi14, Shull+14, glenn15, Burchett15, Lehner16,  Codoreanu18, Rudie19, Dutta20, H20, PH20}.

Studies which target quasar absorption lines arising in the CGM of their host galaxies have revealed metal-bearing gas out to large galactocentric distances for both low-ionization {\MgIIdblt} selected absorbers \citep[e.g.,][]{BB91, SDP94, glenn08, Chen10, nikki13b, MF19, Huang21} and for higher-ionization {\OVIdblt} selected absorbers \citep[e.g.,][]{Tumlinson11, Stocke13, Mathes14, glenn15, Nikki17, Ng19}. Such studies targeting the intermediate ionization gas commonly observed via the {\CIVdblt} doublet, though not as numerous, have shown that this ion can also exist at large galactocentric distances \citep[e.g.,][]{Chen01, Borthakur13, Bordoloi14, LC14, Burchett16, Rudie19, MF21}.

However, to the degree that these ions and their spatial distribution around galaxies have been characterized, it remains unclear (1) how far out within the DM halo (or beyond the DM halo) metals are distributed, (2) what the origins of these metals are, i.e., what percent comes from outside the central galaxy and what percent from inside, and (3) what accretion modes, stellar feedback mechanisms, and recycling processes are distributing metals to such large galactocentric distances.  It is likely that the answers to these questions will change in their details as a function of both redshift and DM halo mass. 

Global insights into the statistical relationship between metal-line absorbing structures at different redshifts, DM halos, and their central luminous galaxies, can be inferred by comparing their cosmic distributions. Over the last decade, tremendous progress has been made in characterizing the DM halo mass function \citep[e.g.,][]{Murray13, Behroozi13,Bocquet16} and galaxy luminosity function \citep[e.g.,][]{Weisz14,Parsa16,Bouwens21} over the redshift ranges corresponding to those for which metal-line absorbers have been characterized.

In \citet[][hereafter \citetalias{H20}]{H20}, we used Keck/HIRES \citep{Vogt94} and VLT/UVES \citep{Dekker00} spectra, including from the KODIAQ \citep{OMeara15} and UVES SQUAD \citep{Murphy19} projects, to expand and extend the survey work of \citet{Cooksey13} and constrained the global statistics of {\CIV} absorbers to a limiting equivalent width detection sensitivity of ${\wrlim} \!=\! 0.05$~{\AA} over the redshift range $1.0 \!\leq\! z \!\leq\! 4.75$.  Using the works of \citet{Cooksey10} and \citet{Burchett16}, we extended our investigation to cover $0 \!\leq\! z \!\leq\! 1$. We found that the cosmic incidence of {\CIV} absorbers increases with time across $0 \!\leq\! z \!\leq\! 5$ (corresponding to the last $\sim$12.5~Gyr, or $\sim$90\% of cosmic time), with the rise being more rapid as the {\EWr} of the subsample is increased. The evolution of {\CIV} absorbers with {\wweak} and {\wstrong} occurred most prominently prior to and throughout the so-called ``Cosmic Noon'' epoch ($1.5 \!\leq\! z \!\leq\! 3$), when cosmic star formation, black hole activity, galactic scale winds, and gas accretion rates reached their peak \citep[e.g.,][]{vdv11b, MD14, Feldman16, Rupke18}.  On the contrary, {\CIV} absorbers with {\wvweak} evolved very little until after Cosmic Noon was completed, at which time their cosmic incidence, $dN/dX$, rose by a factor of $\sim$4.5.   

In this paper, we use the observed evolution of {\CIV} absorber populations in concert with the evolution of galaxy populations to better constrain the statistical extent of metal-enriched gas as a function of galaxy properties and to elucidate the connection between enriched gas, galaxies, and DM halos as manifested by the baryon cycle. In the most general sense, properties of gaseous structures denser than the cosmic mean should be causally linked to galaxies, just as galaxy growth is causally connected to halo growth \citep[e.g.,][]{Kravtsov04,More09,Moster18,Behroozi19,Whitler20}. We exploit this connection to understand how the spatial distribution of metals relative to galaxies changes with galaxy luminosity, halo mass, and redshift.


This paper is organized as follows. In Section~\ref{sec:galsdm}, we connect the populations of galaxies to that of DM halos and investigate the connection between absorbers and halos. We leverage our measurements to estimate the statistical extent of {\CIV} absorbing gas around galaxies of different luminosities in Section~\ref{sec:extent}. We discuss our results in the context of theory and observations and their implications for galaxy evolution and the baryon cycle in Section~\ref{sec:discuss}. 
Finally, the conclusions are presented in Section~\ref{sec:conclude}.
As in \citetalias{H20}, we adopt the WMAP5 cosmology, with $H_{0}=71.9$ {\kmsmpc}, $\Omega_{\hbox{\tiny M}} = 0.258$, and $\Omega_{\Lambda} = 0.742$ \citep{WMAP09}.


\section{Gas, Galaxies, and Halos}
\label{sec:galsdm}

In this section, we investigate the cosmic incidence of DM halos and compare them to the cosmic incidence of {\CIV} populations that we measured in \citetalias{H20}. 
For a given population of objects, the observed cosmic incidence, or ``co-moving path density,'' is proportional to the product of the cosmic number density, $n(z)$, and the physical cross-section, $\sigma(z)$,
\begin{equation} 
\frac{dN}{dX}(z) = \frac{c}{H_0} n(z) \sigma(z) \, .
\label{eq:dndx_abs}
\end{equation}

In general terms, {\dndx} is the number of structures intercepted by a line of sight per unit of ``absorption distance,'' $X(z)$ \citep{BP69}. For absorbing systems, this quantity is interpreted as the number of intervening absorption systems observed per unit co-moving line-of-sight path length. 
Similarly, we can interpret {\dndx} for a population of DM halos as the number of virial halos, defined by $R_{v}(M_h,z)$, the virial radius of a DM halo of mass $M_h$ at redshift $z$, intercepted per unit co-moving path of a given line of sight.
A comparison of the evolution of cosmic incidence of gas absorbers and DM halos provides insight into the spatial distribution of gas structures in the universe.

To examine the cosmic incidence of {\CIV} absorbers relative to DM halos, we define the ``relative gas incidence,''
\begin{align} 
{\fX}(z) = 
\frac
{\displaystyle 
\frac{dN}{dX}(z)\Big|_{\hbox{\small {\CIV}}(\EWr \geq \wrlim)}
}
{\displaystyle 
\frac{dN}{dX}(z)\Big|_{\hbox{\small \Rv}(M_h \geq {\Mhmin}(z))}
} 
\, ,
\label{eq:fX}
\end{align}
as the ratio of co-moving path density of {\CIV} absorbers with $\EWr \!\geq\! \wrlim$ to that of spherical DM halos with ${M_h \!\geq\! {\Mhmin}(z)}$.  

Under the assumption that {\CIV} absorbers are co-spatial with DM halos, the relative gas incidence quantifies the relative cross-sectional size of {\CIV} absorbers with $\EWr \!\geq\! \wrlim$ with respect to the halo mass-weighted cross-section of DM halos with $M_h \!\geq\! {\Mhmin}(z)$.  A value of ${\fX}(z) \!<\! 1$ implies that the sizes of ``{\CIV} halos'' with $\EWr \!\geq\! \wrlim$ are smaller than the halo mass-weighted virial radii of DM halos, thus suggesting that such {\CIV} absorbers, on average, spatially reside within the virial radii of DM halos.  A value of ${\fX}(z)\!>\!1$ implies that the sizes of ``{\CIV} halos'' with $\EWr \!\geq\! \wrlim$ are larger than the halo mass-weighted virial radii of DM halos, thus suggesting that such {\CIV} absorbers, on average, reside outside the virial radii of DM halos. This assumption of equal abundance of {\CIV} absorbers and DM halos is quite restrictive and is likely to be incorrect, if for instance, there are significant numbers of satellite galaxies surrounding massive centrals in a halo \citep[e.g.,][]{Zheng05,Mandelbaum06}. We discuss this assumption in section~\ref{sec:discuss} and elsewhere below.

If instead, we assume that absorbing halos and DM halos have the same physical extent, {\fX} measures the relative abundance of {\CIV} absorbers relative to DM halos. In this scenario, ${\fX}(z)\!>\!1$ implies that absorbers with $\EWr \!\geq\! \wrlim$ are more abundant than DM halos with  ${M_h \!\geq\! {\Mhmin}(z)}$ and ${\fX}(z)\!<\!1$ implies the converse. 
Without either of the above assumptions, the relative gas incidence quantifies the product of abundance and cross-section of absorbers relative to that of DM halos.

The numerator of Eq.~\ref{eq:fX} has been measured directly from the data,  survey path length, and completeness functions for ${\wrlim} \!=\! 0.05, 0.3$, and $0.6$~{\AA} as described in \citetalias{H20}. There are no assumptions required for the calculation, for example, about the gas physical conditions, cosmic number densities, or absorber cross-sections, as it is simply an exercise in counting absorbers. The computation of the denominator, on the other hand, relies on our cosmic census of DM halos and their evolution. 
We computed the denominator of Eq.~\ref{eq:fX} from Eq.~\ref{eq:dndx_abs} using
\begin{equation} 
n(z) \sigma(z) = \int_{M_{h,\mathrm{min}}}^{\infty} \!\!\!\!\!\!\!\!\!\!
\phi(M_h,z) \, \pi R^2_{v}(M_h,z) \, dM_h \, ,
\label{eq:nsighalo}
\end{equation}
where $\phi(M_h,z)$ is the DM halo mass function (number density of DM halos with mass $M_h$  per unit halo mass) at redshift $z$, and ${\sigma(M_h,z) = \pi R^2_{v}(M_h,z)}$ is the assumed spherical cross-sectional area.

\begin{figure}[htbp]
\vglue 0.1in
\includegraphics[width=0.45\textwidth]{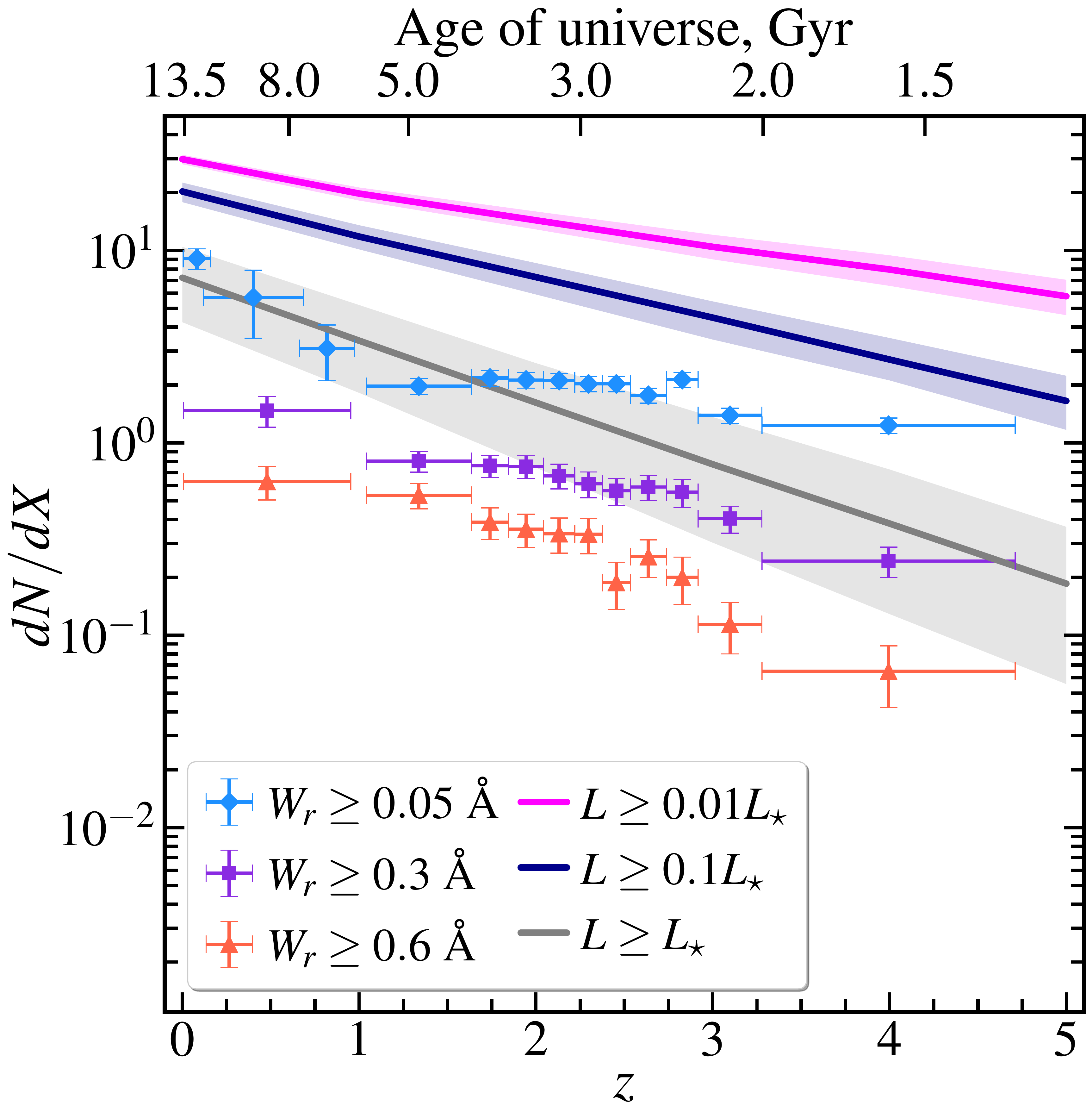} \vspace{-5pt}
\caption{Comparison of cosmic incidence of {\CIV} absorbers and DM halos. The measured {\dndx} of different populations of {\CIV} absorbers are shown by light blue ({\wvweak}), purple ({\wweak}), and orange ({\wstrong}) data points; the uncertainties are $\pm1\sigma$. The computed {\dndx} of halos (Eq.~\ref{eq:nsighalo}) are represented by solid curves; halos hosting galaxies with ${L\geq{\onepLstar}}$ are shown in pink, ${L\geq{\tenpLstar}}$ is dark blue, and ${L\geq{\Lstar}}$ is grey. 
The shading on these curves reflects the uncertainty in the halo mass obtained from abundance matching to luminosity (see Appendix~\ref{app:abundance}).
} 
\label{fig:dndx_halo+abs}
\end{figure}

Historically, the sizes of absorbing structures tracing various ionization phases of gas have been linked to the directly-observed luminosity of galaxies around which they are found \citep[e.g.,][]{BB91, Lanzetta93, Steidel95, Chen01,Chen10,glenn08,Ribaudo11,nikki13b}. In order to map Eq.~\ref{eq:nsighalo} to a minimum central galaxy luminosity, {\Lmin}, we apply the technique of abundance matching \citep[e.g.,][]{Kravtsov04,TG11,Behroozi13,Behroozi19}, thus mapping the cumulative number density of halos with $M_h \!\geq\! {\Mhmin}$ at redshift  $z$ to that of galaxies with $L \!\geq\! {\Lmin}$.  We adopted the 1500~{\AA} UV luminosity functions of \citet{Parsa16} and the DM halo mass functions of \citet{Behroozi13} across $0 \!\leq\! z \!\leq\! 5$. Details on how we perform abundance matching and compute Eq.~\ref{eq:nsighalo} are presented in Appendix~\ref{app:abundance}.

\begin{figure*}[htbp]
\gridline{\fig{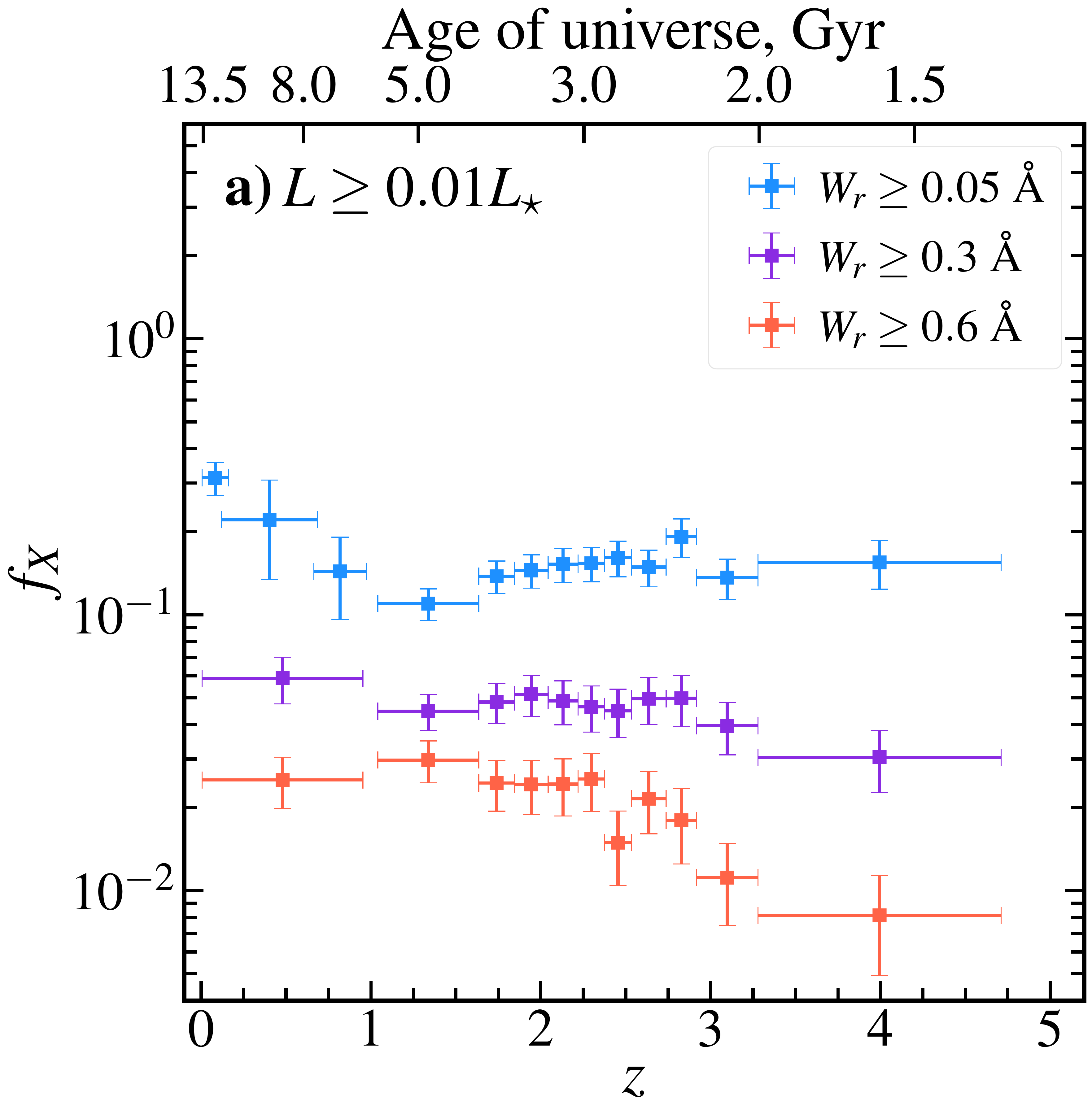}{0.34\textwidth}{}
 \fig{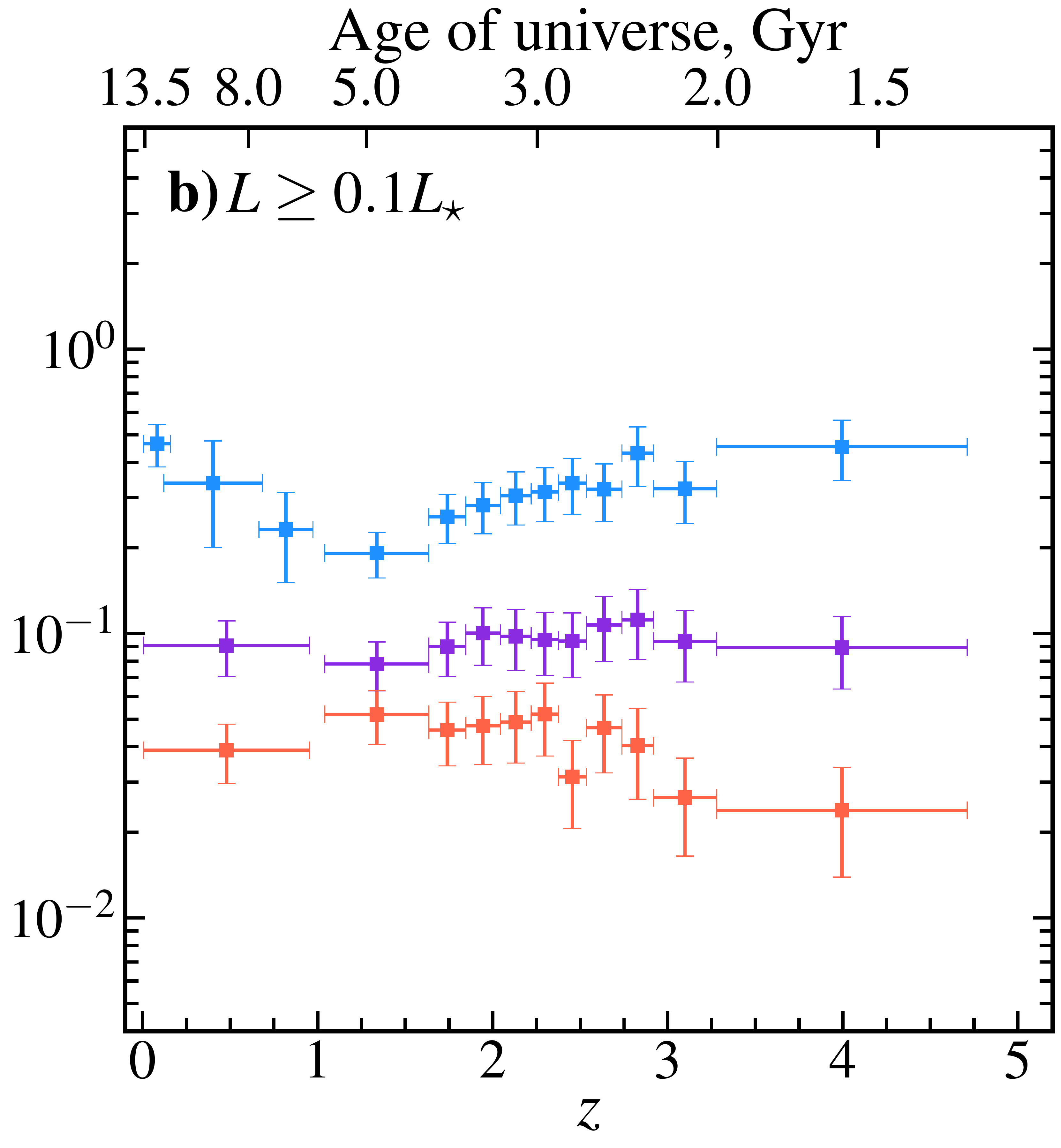}{0.32\textwidth}{}
 \fig{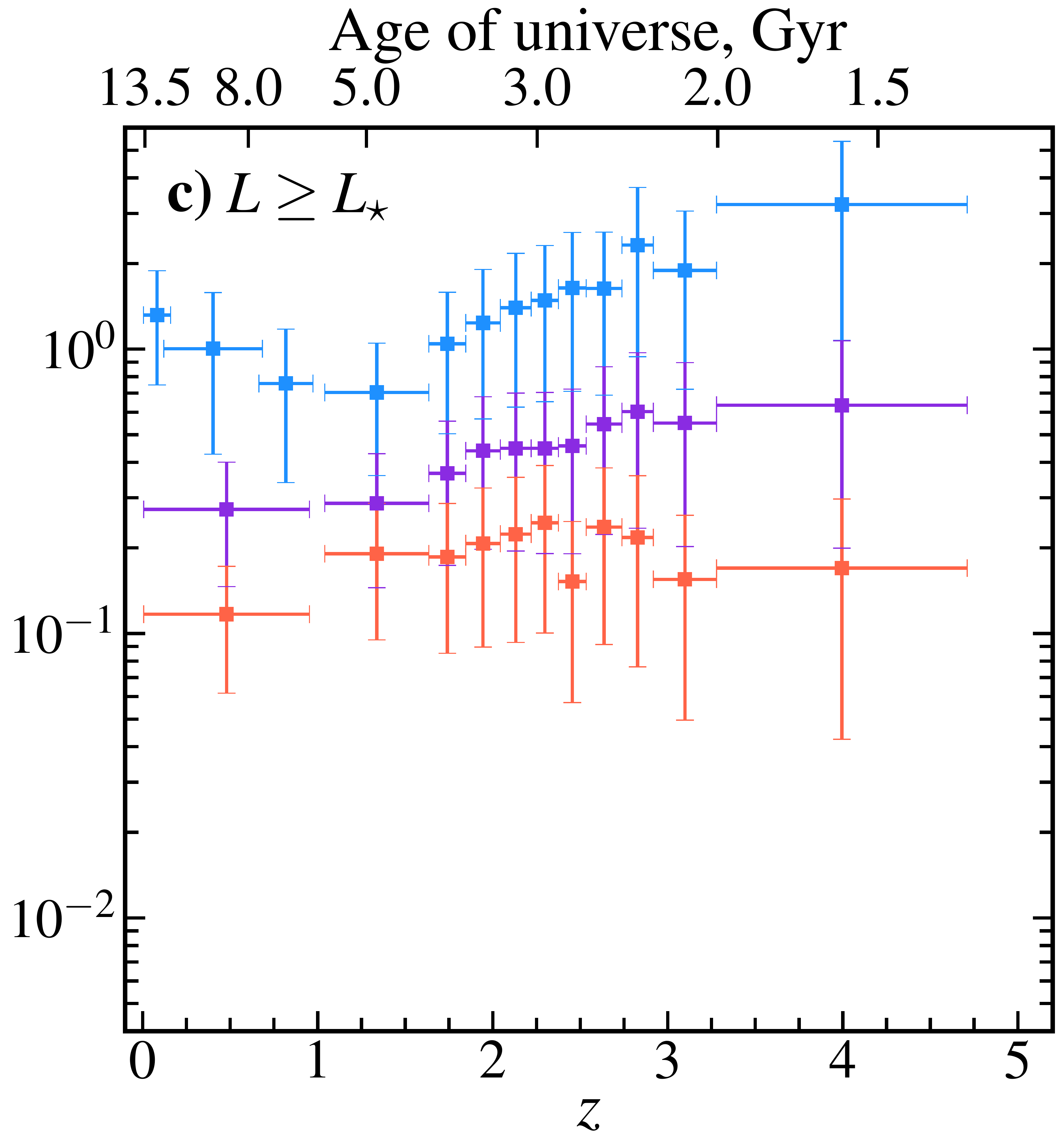}{0.32\textwidth}{}
        } \vspace{-25pt}
\caption{
The {\CIV} absorption to DM virial halo ratios (Eq.~\ref{eq:fX}) as functions of redshift for halos with (a) $L\!\geq\!{\onepLstar}$, (b) $L\!\geq\!{\tenpLstar}$, and (c) $L\!\geq\!{\Lstar}$. The blue, purple and orange data represent ${\fX}$ for {\CIV} gas with {\wvweak}, {\wweak}, and {\wstrong}, respectively. 
}
\label{fig:fX}
\end{figure*}

In Figure~\ref{fig:dndx_halo+abs}, we compare the measured {\dndx} for {\CIV} absorbers with the thresholds ${\wrlim}=0.05$, 0.3, and 0.6~{\AA} \citepalias{H20} to the {\dndx} of DM halos with ${\Mhmin}(z)$ corresponding to fixed ${\Lmin}\!=\! {\onepLstar}$, {\tenpLstar}, and {\Lstar}. The shaded bands represent the uncertainty in luminosity obtained from abundance matching.
We added a data point for the {\dndx} of {\wweak} absorbers at $\langle z \rangle \!\simeq\! 0.5$, from \citet{Cooksey10}, which was not presented in \citetalias{H20}.

Our choice of ${\Lmin}\!=\!{\onepLstar}$ follows from the COS-Dwarfs survey, which found {\CIV} around galaxies with UV $1500$~{\AA} luminosities $L\!\sim\!0.005$--0.14{\Lstar} at $z \!\leq\! 0.1$ \citep{Bordoloi14}. We choose ${\Lmin}\!=\!{\tenpLstar}$ as one of the luminosity cutoffs based on the findings of \citet{Burchett16} that at $z\!\simeq\!0$, the {\CIV} covering fraction within 1~{\Rv} declines below 10\% for galaxies with stellar mass below {$\log M_{\star} \!\sim\! 9.5$~{\Msun}}. This translates to $M_h \!\approx\! 3\times10^{11}$~{\Msun} according to the stellar-halo mass relation of \citet{Girelli20}, and is roughly the halo mass for a {\tenpLstar} galaxy at $z \!\simeq\! 0$ (see Figure~\ref{fig:hmlr}). Finally, we choose the ${\Lmin}\!=\!{\Lstar}$ to examine the most massive galaxies at each redshift.

The results in Figure~\ref{fig:dndx_halo+abs} indicate that the cumulative co-moving path density of halos hosting $L \!\geq\! {\onepLstar}$ and ${L \!\geq\! {\tenpLstar}}$ galaxies exceeds that of {\CIV} absorbers at all redshifts for thresholds of ${\wrlim}\!=\!0.05$~{\AA} and above. At $z \!>\! 2$, the $L \!\geq\! {\Lstar}$ halo {\dndx} is smaller than the {\wvweak} {\CIV} {\dndx}, though they roughly trace each other for $z \!<\! 2$. Our exercise would suggest that the product of the number density and cross-section, $n(z)\sigma(z)$, of the weakest {\CIV} absorbers exceeds that of the most massive halos at early epochs.

Applying Eq.~\ref{eq:fX}, we then compute the gas incidence ${\fX}(z)$ of {\CIV} absorbers relative to the virial radius as a function of redshift for our six combinations of ${\wrlim}\!=\!0.05$, $0.3$ and 0.6~{\AA} and ${\Lmin}\!=\!0.01{\Lstar}$, $0.1{\Lstar}$, and {\Lstar} and present them in Figure~\ref{fig:fX}. Each panel shows a different {\Lmin} halo population.

For {\CIV} absorbers with {\wvweak} we find ${\fX} \!\simeq\! 0.1$--$0.5$ for halos hosting galaxies with $L\geq {\onepLstar}$ and {\tenpLstar}. For $L\!\geq\!{\Lstar}$, we find ${\fX}\!\!\simeq\! 0.8$--$3$. The value of ${\fX} \!>\! 1$ for $L\!\geq\!{\Lstar}$ halos at $z\!=\!0$ and $z\!\geq \!2$ means that all the observed {\wvweak} absorbers cannot be accounted for by massive halos with $L\!\geq\!{\Lstar}$ galaxies at those redshifts. 
If {\wvweak} {\CIV} absorbers are spatially coincident with $L\!\geq\!{\Lstar}$ halos, ${\fX} \!>\! 1$ implies that the absorbing structures extend beyond the virial radii of the halos, likely in the IGM between halos. If {\wvweak} {\CIV} absorbing structures have the same physical extent as $L\!\geq\!{\Lstar}$ virial halos, then {\wvweak} {\CIV} is more abundant than $L\!\geq\!{\Lstar}$ halos, again implying that some {\CIV} absorbing systems must live in the IGM. We can also conclude that $L\!\geq\!{\onepLstar}$ and $L\!\geq\!{\tenpLstar}$ halos are more abundant and/or larger in physical size than {\wvweak} {\CIV} absorbing structures.

At early epochs, from $z\!=\!4$ to $z\!=\!2$, the evolution of {\fX} for {\CIV} absorbers with {\wvweak} declines by a factor of three for the most massive halos. This evolution is less pronounced for halos hosting $L\!\geq\! {\tenpLstar}$ galaxies, and virtually absent for $L\!\geq\!{\onepLstar}$ halos. This behavior would suggest that, at early epochs, the product $n(z)\sigma(z)$ of the most massive halos grows at a rate that exceeds that of {\wvweak} {\CIV} absorbing structures, but that this difference in evolution diminishes as lower mass halos are included in the halo population.

What is interesting for {\wvweak} {\CIV} absorbers is that their {\it relative\/} gas incidence has a minimum at $z\!\simeq\! 1.5$ followed by an increase toward the present epoch, $z\!=\!0$; this evolution is observed for all halo masses. There are two noteworthy inferences that can be drawn from this behavior. First, this may suggest that the spatial extent of weak {\CIV} absorbers in the outer regions of galactic halos is decreasing during the epoch of Cosmic Noon, when the global star formation is at its peak \citep{MD14}. Second, following Cosmic Noon, the typical physical extent of {\wvweak} {\CIV} absorbing structures likely grows faster than the virial radius of halos.

For stronger {\CIV} absorbers, {\wweak} and {\wstrong}, we find ${\fX}\!<\!1$ for all halo masses. This may suggest that higher column density (and/or kinematically complex) {\CIV} absorbing structures reside within the virial radii of galaxy halos of all masses. It may also suggest that all halo populations are significantly more abundant than stronger absorbers.

For {\wweak} and {\wstrong} absorbers, the early epoch evolution of relative gas incidence becomes most pronounced as lower mass halos are included in the halo population. For $L\!\geq\!{\onepLstar}$ galaxies, ${\fX}$ for {\wstrong} {\CIV} absorbers increases by a factor of a few from $z\!=\!4$ to $z\!=\!2$ and then flattens out.  Thus, when the lowest mass halos are included, the {\wstrong} {\CIV} absorbing structures are growing in size and/or number density at a rate that exceeds that of galaxy halos prior to Cosmic Noon, but then after Cosmic Noon, the halo and absorber sizes remain in lock step. Similar behavior is found for {\wweak} {\CIV} absorbers, though the early epoch evolution is less pronounced. When we limit our consideration to $L\!\geq\!{\tenpLstar}$ halos, we find that the early epoch evolution is less pronounced.

The takeaways from Figure~\ref{fig:fX} are that 
(a) $L\!\geq\!{\Lstar}$ galaxy halos cannot account for all the observed {\wvweak} {\CIV} absorbers, but $L\!\geq\!{\onepLstar}$ and $L\!\geq\!{\tenpLstar}$ halos can,
(b) if {\CIV} absorbers are co-spatial with DM halos, then all absorbers should live inside the virial radius of $L\!\geq\!{\onepLstar}$ and $L\!\geq\!{\tenpLstar}$ halos, but {\wvweak} absorbers may live beyond the virial radius of $L\!\geq\!{\Lstar}$ halos,
(c) as lower mass halos (lower luminosity galaxies) are accounted for, the physical extent of {\CIV} absorbing gas resides progressively in the inner regions of a halo and/or absorbers become much less common than halos, and
(d) for {\wvweak} absorbers, there is a minimum in the incidence of the {\CIV} absorbing structures relative to virial halos at $z\!\simeq\!1.5$, shortly after Cosmic Noon, followed by a remarkable relative increase at $z\!\leq\!1$.

As informative as these concluding remarks may be, some of them require key assumptions about the association of {\CIV} absorbers and galaxy halos. The assumption that each individual {\CIV} absorbing structure is coincident with a single DM halo hosting a single galaxy may not be an adequate or accurate description of absorber-galaxy associations, as we discuss later. Regardless, the analysis above does provide insights into the relative global evolution of the product of cosmic number density and cross-sectional area of cumulative populations of {\CIV} absorbers and galaxy halos.


\section{The extent of {\CIV} around galaxies}
\label{sec:extent}

In this section we aim to characterize the redshift evolution of the ``gas radius'' as a function of galaxy luminosity, which we refer to as $R_g({\LLstar},z)$. We first describe how we estimate the gas radius, then we examine its evolution for a fixed galaxy luminosity. Finally, we examine the redshift evolution as a function of galaxy luminosity. Our focus is to obtain some insight on how $R_g({\LLstar},z)$ compares and evolves relative to the virial radii of halos hosting galaxies of various luminosities.

\subsection{Estimating the Gas Radius}

\begin{figure*}
\gridline{\fig{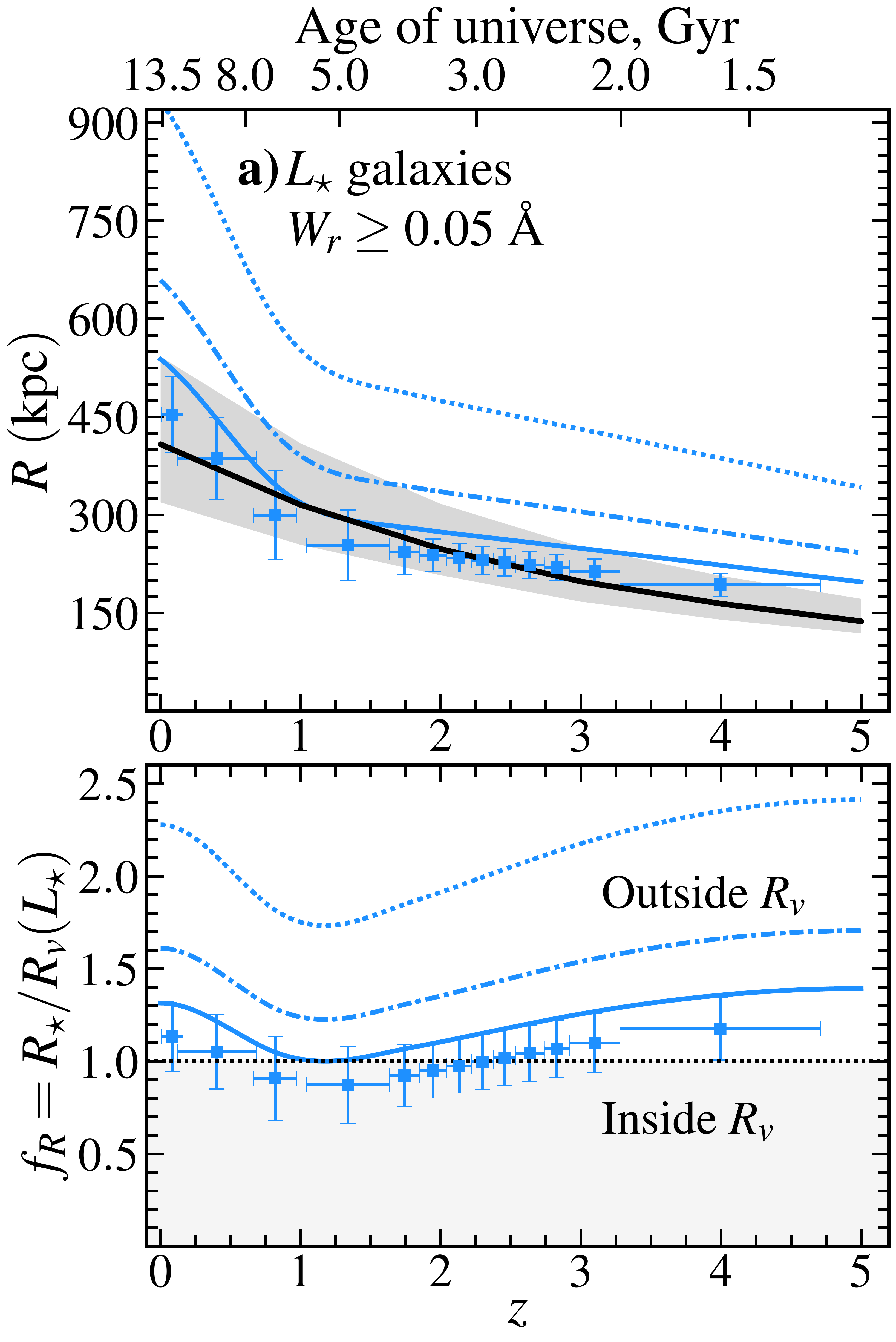}{0.34\textwidth}{}
 \fig{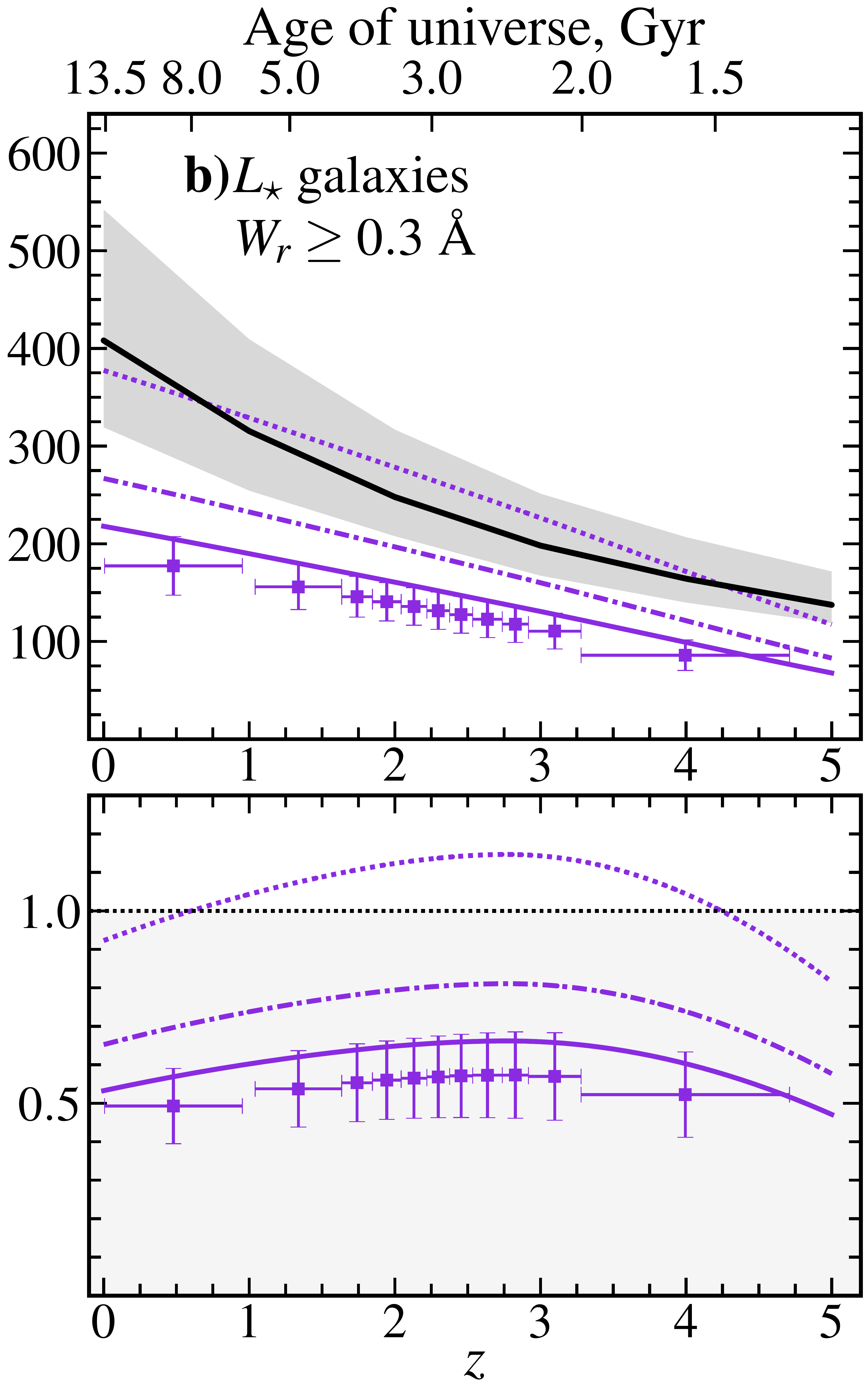}{0.315\textwidth}{} \fig{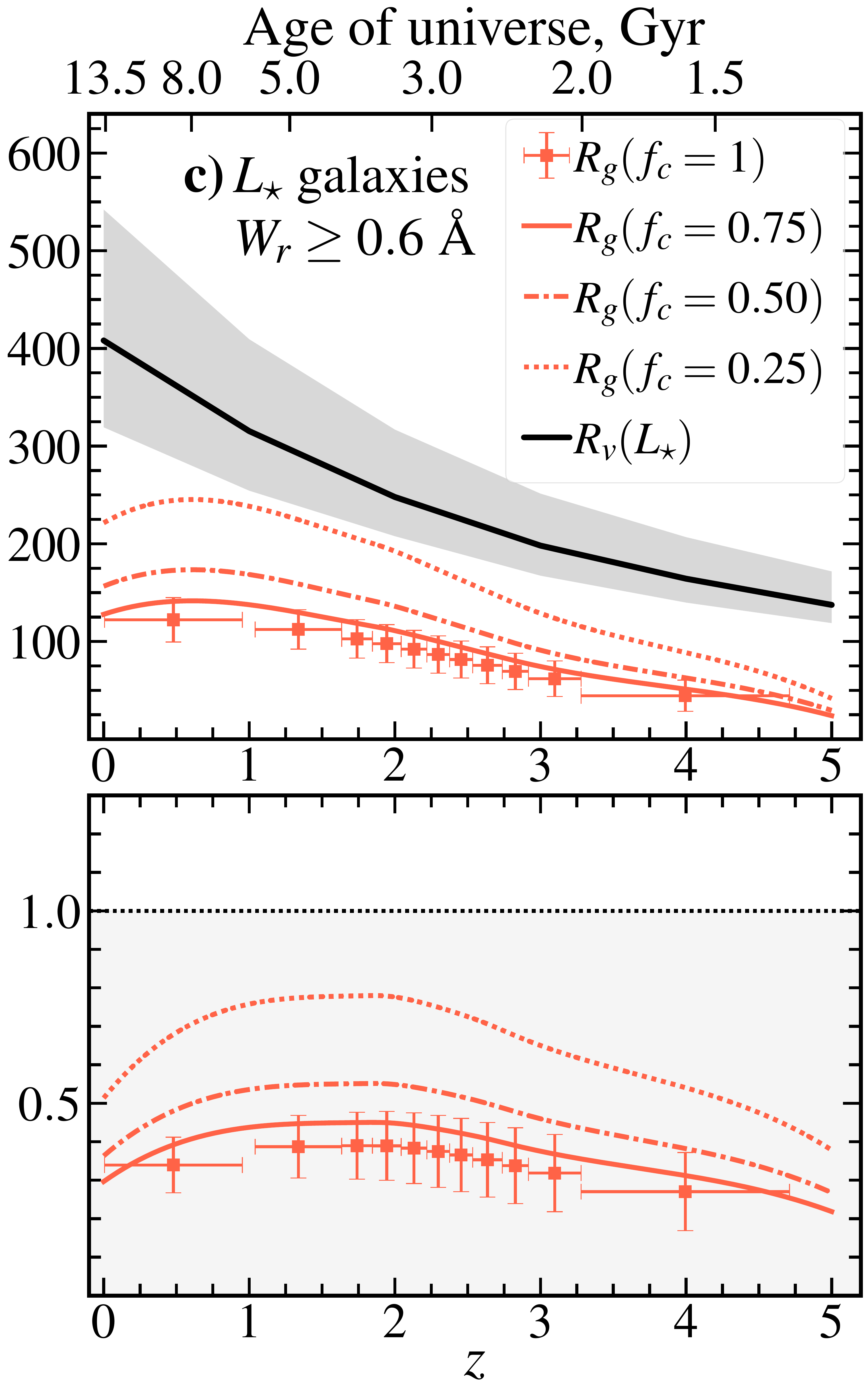}{0.315\textwidth}{}
        } \vspace{-25pt}
\caption{
Evolution of characteristic {\CIV} gas radius {\Rstar} (Eq.~\ref{eq:rstar}), and virial radius {\Rv} (Eq.~\ref{eq:rvir}), for {\Lstar} galaxies, assuming galaxies with ${\Lmin} \!=\!{\onepLstar}$ are associated with the absorbers. {\it Top panels}: $R_g\!=\!{\Rstar}$ as a function of redshift are shown for (a) ${\wrlim}\!=\!0.05$~{\AA} absorbers (blue data points and curves); (b) ${\wrlim}\!=\!0.3$~{\AA} absorbers (purple data points and curves); and (c) ${\wrlim}\!=\!0.6$~{\AA} (orange data points and curves). The black curves and grey shading represent {\Rv}({\Lstar}) and $\pm1\sigma$ uncertainties as a function of redshift. {\it Bottom panels}: $f_R \!=\! {\Rstar}/{\Rv}({\Lstar})$ shown as a function of redshift for each population. In each panel, the data points and error bars represent the value obtained when $f_c=1$ is assumed, while the solid, dash-dotted, and dotted curves show the values obtained if we assume $f_c\!=\!0.75$, $0.50$, and $0.25$, respectively. The horizontal dotted black line shows $f_R\!=\!1$, above which absorbers live outside the DM halo. We adopt $\beta\!=\!0.5$ based on \citet{Chen01}.
}
\label{fig:lstar}
\end{figure*}

Under the assumption that each {\CIV} absorber is associated with a single galaxy halo, the cosmic number density of gas structures, $n(z)$ in Eq.~\ref{eq:dndx_abs}, is given by the number density of galaxies above a minimum luminosity, {\Lmin}. The cross-section of absorbing gas is $\sigma(z) = \pi R_g^2({\LLstar},z)$, where the characteristic gas absorption radius for a given {\wrlim} is assumed to scale with galaxy luminosity as
\begin{equation} 
R_g ({\LLstar},z) = R_{\star}(z) ~({\LLstar})^{\beta} \, ,
\label{eq:rg}
\end{equation}
where ${\Rstar}(z)$ is the characteristic absorption radius for an {\Lstar} galaxy at redshift $z$, and $\beta$ is the Holmberg scaling parameter \citep[e.g.,][]{holmberg75}, which is commonly employed to parameterize how the gas radius scales with galaxy luminosity \citep[e.g.,][]{BB91, Steidel95, Chen01, glenn08, nikki13b}.  

To compute ${\Rstar}(z)$, we use the method employed in \citetalias{H20} using the observed {\CIV} incidence, which we summarize in Appendix~\ref{app:rstar}. Importantly, ${\Rstar}(z)$ is not a function of {\dndx} and $z$ only, but also of the assumed minimum galaxy luminosity ${\Lmin}/{\Lstar}$, the Holmberg power-law luminosity scaling $\beta$, the multi-parameter luminosity function, $\phi(\phi_{\star},{\LLstar},\alpha,z)$, and the absorbing gas covering fraction $f_c(z)$.

With regard to the covering fraction, sparse data are available for {\CIV} absorbers. At $z\!\sim\!0$, \citet{Bordoloi14} measured the covering fraction within $\sim$0.5{\Rv} of galaxies with $L\!\sim$0.005--0.14{\Lstar} to be $\langle f_c\rangle \!\simeq\! 0.5$, while \citet{Burchett16} found $\langle f_c\rangle \!\simeq\! 0.33$ within {\Rv} of $L\!\geq \!{\onepLstar}$ galaxies. From stacked spectra of $L\!\geq\! {\onepLstar}$ galaxies, \citet{LC14} estimated a mean within {\Rv} of $\langle f_c\rangle \!\simeq\! 0.45$ at $z\!<\!0.18$. These studies all had a detection threshold of ${\wrlim} \!\simeq\! 0.05$~{\AA}.
\citet{MF21} constrained the covering fraction of {\wvweak} absorbers at $z\!\sim$1--1.5 around $L\!\geq\! {\onepLstar}$ galaxies. 
Based on their model for {\CIV} covering fraction, we estimate $\langle f_c\rangle \!\simeq\! 0.25$ within {\Rv}.
Finally, for $z\!\sim\!2\!-\!3$, \citet{Rudie19} estimated the covering fraction of {\CIV} around eight $L\!\geq\! {\tenpLstar}$ galaxies. Adopting their column density threshold of $\log N/\mathrm{cm}^{-2} \!=\! 13$ as the rough equivalent of ${\wrlim}\!=\!0.05$~{\AA}, we estimate the average of their measurements to be $\langle f_c\rangle \!\simeq\! 0.75$ within 100 kpc. 
As for the uncertainty in the covering fraction, ${\Rstar}(z)$ scales as $f_c^{-1/2}(z)$. Our fiducial calculations employ $f_c(z)\!=\!1$, but we also explore the values $f_c(z) \!=\! 0.25$, 0.50, 0.75, and assume no redshift evolution.

We assume ${\Lmin}\!=\!{\onepLstar}$ based on both \citet{Bordoloi14} and \citet{Burchett16} finding absorption around $L \!\geq\! {\onepLstar}$ galaxies. For these plots, we assume $\beta\!=\!0.5$ following the findings of \citet{Chen01}.
Note that $\beta$ could be a function of $z$, ${\LLstar}$, and/or {\wrlim}. As there are no observational constraints on such dependencies with $\beta$, we assume it is a single valued constant. For {\CIV} absorbers with ${\wrlim}\!\simeq\!0.1$~{\AA}, \citet{Chen01} reports $\beta \!=\! 0.5$ for $B$-band luminosity at $z\!\leq\!1$. We use $\beta\!=\!0.5$ for our fiducial calculations, however, we explore a range of values between $0 \!\leq\! \beta \!\leq\! 0.5$. 
Once we compute ${\Rstar}(z)$, we can obtain $R_g({\LLstar},z)$ from Eq.~\ref{eq:rg} as a function of galaxy luminosity for the assumed value of $\beta$.

\subsection{Evolution of the Gas Radius}
\label{sec:gasevolve}

\subsubsection{Redshift Evolution for $\beta=0.5$}

In Figure~\ref{fig:lstar} (upper panels), we present the redshift evolution of the gas radius around {\Lstar} galaxies for ${\wrlim}\!=\!0.05$, 0.3, and 0.6~{\AA} {\CIV} absorbers, respectively. In this special case the gas radius given by Eq.~\ref{eq:rg} is $\Rstar(z)$.  We also show the virial radius for {\Lstar} galaxies, $R_v({\Lstar})$, as the black curve with gray shading. The virial radius $R_{v}({\Lstar})$ is calculated from Eq.~\ref{eq:rvir} with $M_h(z)$ corresponding to ${\Lstar}$ using abundance matching at redshift $z$ (see Appendix~\ref{app:abundance}). Recall that the halo mass of an {\Lstar} galaxy evolves as shown in Figure~\ref{fig:hmlr}(b). The characteristic gas radius ${\Rstar}(z)$ for a covering fraction of $f_c(z)\!=\!1$ is given by the data points as computed directly from $dN/dX(z)$ using  Eq.~\ref{eq:rstar} with $\beta\!=\!0.5$. We show the sensitivity to the assumed covering fractions by also presenting $f_c(z)\!=\!0.75$ (dash-dotted curves), $f_c(z)\!=\!0.50$ (solid curves), and $f_c(z)\!=\!0.25$ (dotted curves).

In Figure~\ref{fig:lstar} (lower panels), we plot the ratio of the absorbing gas radius of absorbers to the virial radius of the host galaxy's DM halo, which we will refer to as the ``relative gas radius,''
\begin{equation} 
f_R({\LLstar},z) = \frac{R_g({\LLstar},z)}{R_{v}({\LLstar},z)} \, ,
\label{eq:fR}
\end{equation}
which quantifies the extent of {\CIV}-absorbing gas relative to the virial radius as a function of redshift.

We find that, for {\Lstar} galaxies, absorbers with {\wstrong} reside well within the virial radius at all redshifts, independent of the assumed covering fraction. The relative gas radius appears to slightly increase from high redshift and peak across Cosmic Noon, following which it then decreases toward the current epoch. For {\wweak} absorbers, the relative gas radius is well within the virial radius at all redshifts for $f_c \!>\! 0.25$ and is suggestive of a mild decrease with redshift from $z\!=\!3$ toward the current epoch. For {\wvweak} absorbers, the gas extends out to beyond the virial radius (especially at $z\!\simeq\! 0$ and $z\!\geq\!3$) for $f_c \!\leq\! 0.75$. Interestingly, for these weakest absorbers, the relative gas radius declines from $z\!=\!4$, minimizes across Cosmic Noon, and then increases toward the current epoch.

\begin{figure*}
\gridline{\fig{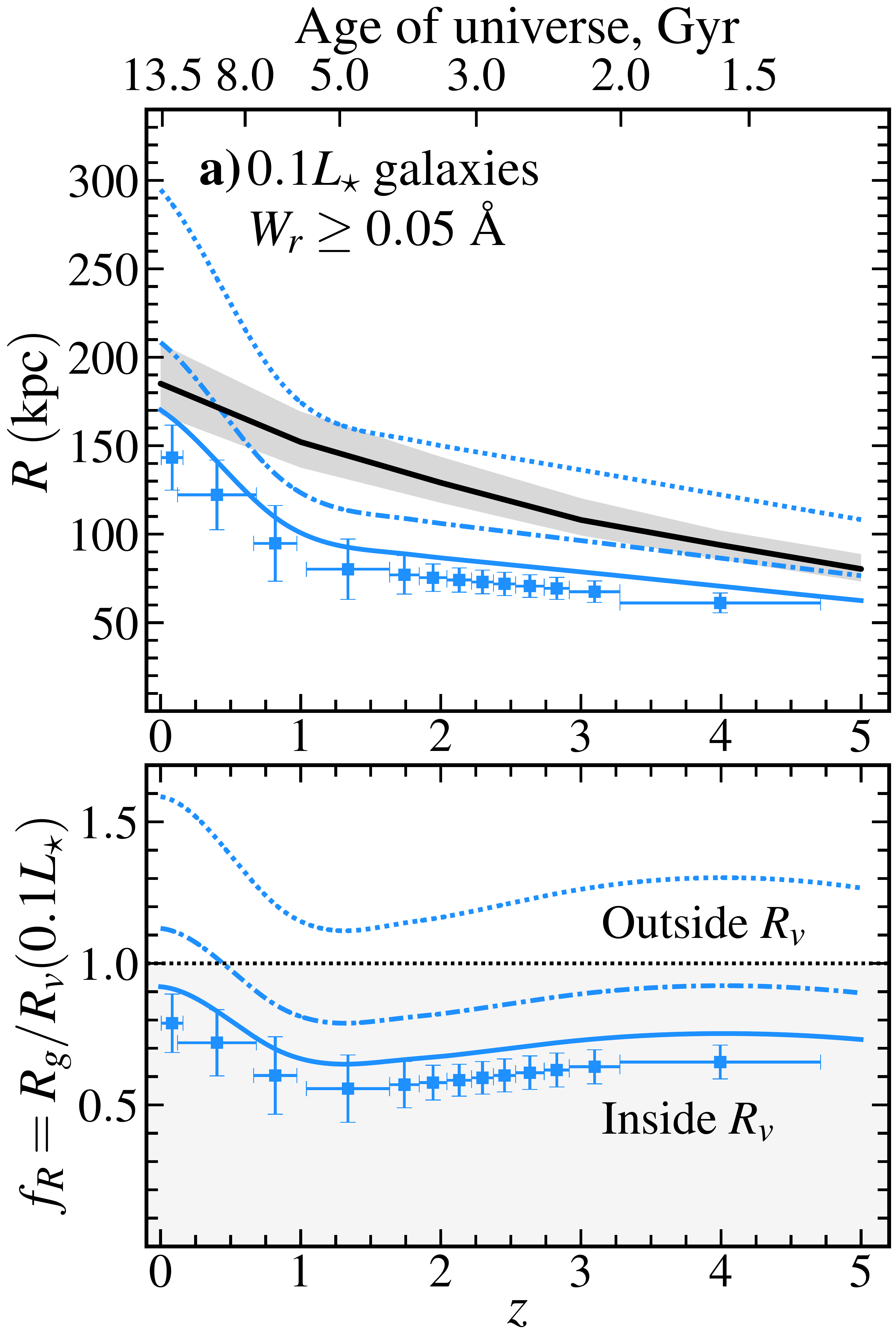}{0.34\textwidth}{}
\fig{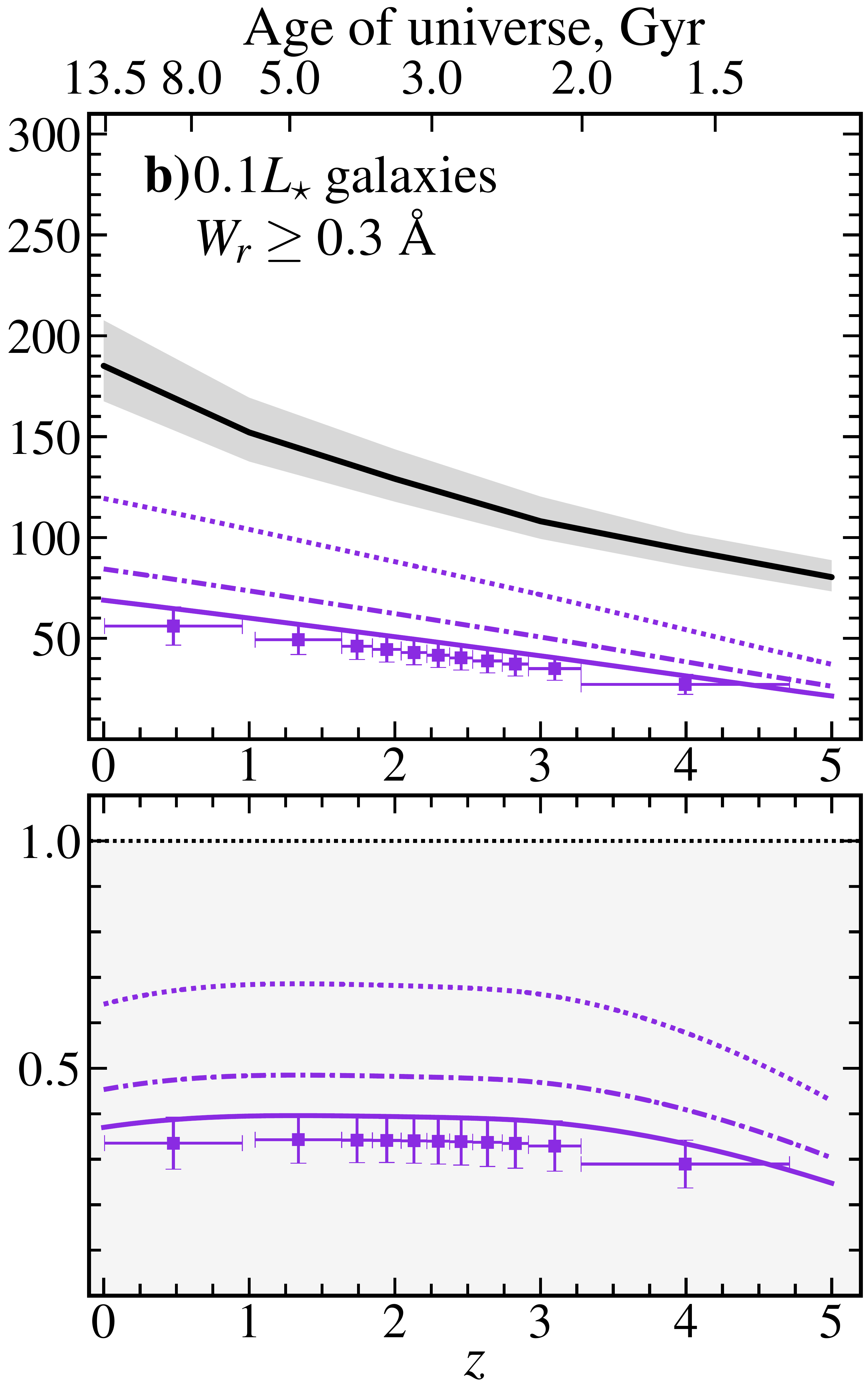}{0.315\textwidth}{}
\fig{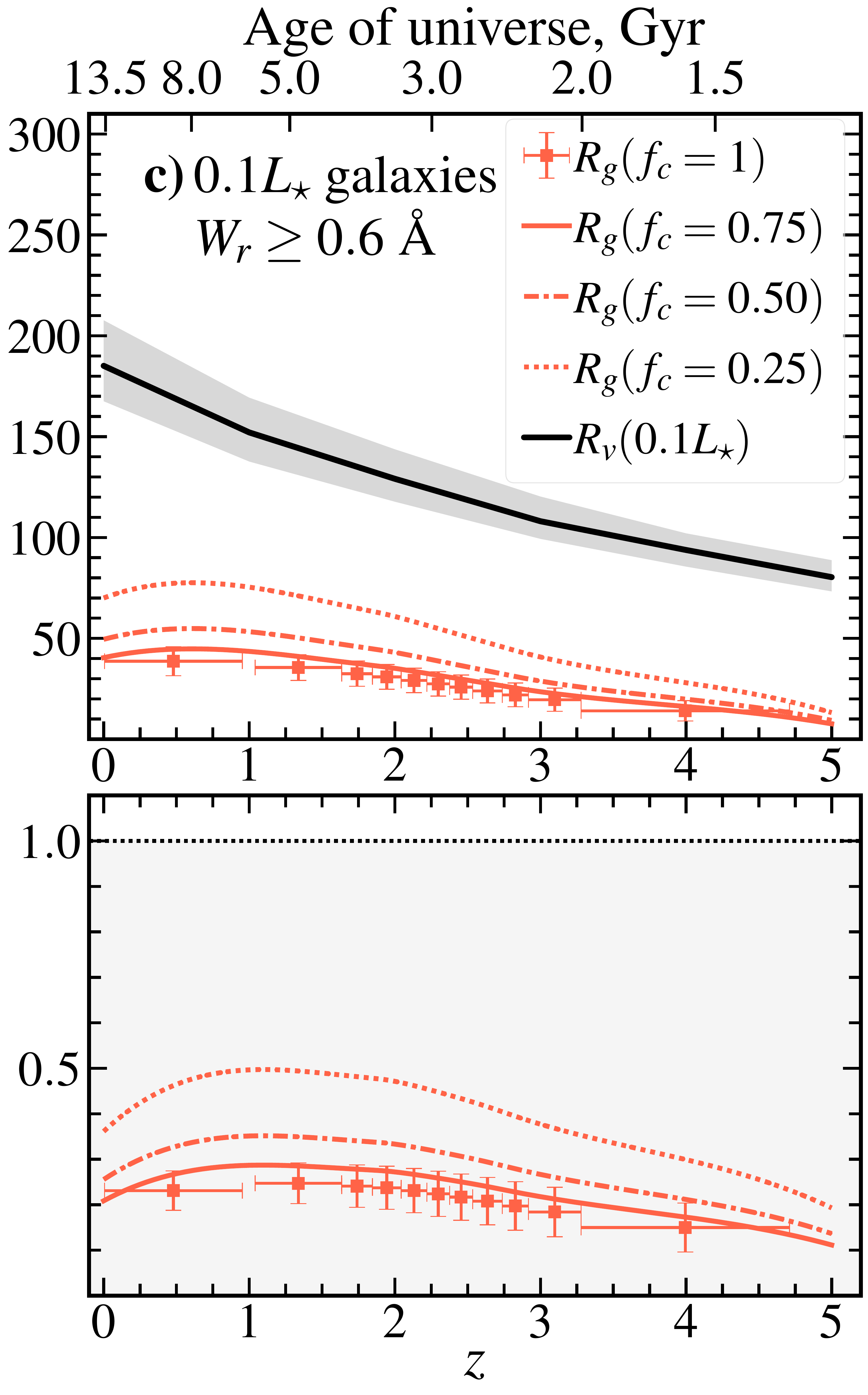}{0.315\textwidth}{}
} \vspace{-25pt}
\caption{
Same as Figure~\ref{fig:lstar}, but for {\tenpLstar} galaxies.
}
\label{fig:10lstar}
\end{figure*}

Using Eq.~\ref{eq:rg} to estimate $R_g({\LLstar},z)$, we can examine $f_R({\LLstar},z)$ for arbitrary {\LLstar}.  As an example, in Figure~\ref{fig:10lstar} (upper panels), we present the redshift evolution of the gas radius for {\tenpLstar} galaxies. The virial radius $R_v({\tenpLstar})$ is shown as the black curve with gray shading.  For $\beta \!=\! 0.5$, we have $R_g({\tenpLstar},z) \!\approx\! 0.32 R_g({\Lstar},z) \!=\! 0.32{\Rstar}(z)$. In Figure~\ref{fig:10lstar} (lower panels), we show $f_R({\tenpLstar},z)$ for ${\wrlim}\!=\!0.05$, 0.3, and 0.6~{\AA} absorbers, respectively. Note that the redshift evolutionary behavior of the relative gas radius follows similar trends as for {\Lstar} galaxies; this is because the gas radius depends on the $dN/dX$ of absorbers and the growth of the virial radius is highly similar for {\tenpLstar} galaxies. However, $f_R({\tenpLstar},z)$ is systematically smaller than $f_R({\Lstar},z)$, suggesting that the extent of {\CIV} absorbing gas relative to the virial radius is smaller in lower luminosity galaxy halos. 
In fact, for {\wvweak} absorbers, the gas radius may not quite extend out to the virial radius, instead only extending beyond {\Rv} for $f_c \!\geq\! 0.25$.

Our analysis would suggest that for {\Lstar} galaxies, {\wweak} {\CIV} absorption typically resides in the inner 50--80\% of the virial halo, and {\wstrong} absorption in the inner 30--50\% of the virial halos for covering fractions of 0.5 or greater. For {\tenpLstar} galaxies, these radii are slightly smaller, 30--50\% and 10--30\% for {\wweak} and {\wstrong} absorption, respectively. 
However, for {\wvweak} absorption, the gas radius may extend, on average, beyond the virial radius for {\Lstar} galaxies and between 60\% and 100\% {\Rv} for {\tenpLstar} galaxies.

\subsubsection{Luminosity Dependent Evolution}

We now examine ${\fR}({\LLstar},z)$ as a function of luminosity and redshift for ${\wrlim}\!=\!0.05$, 0.3, and 0.6~{\AA} {\CIV} absorbers for three different values of $\beta$. Our fiducial value of $\beta \!=\! 0.5$ is motivated by the findings of \citet{Chen01} for ${\wrlim}\!=\!0.1$~{\AA} {\CIV} absorbers at $z\leq1$. This value represents a strong luminosity dependence for the gas radius $R_g({\LLstar},z)$ and will yield smaller gas radii around lower luminosity galaxies. It is not clear if such a strong dependence holds for all galaxy luminosities, redshifts, or ${\wrlim}$. In the absence of constraining data, we adopt a simple bracketing of the luminosity scaling by exploring $\beta\!=\!0$, 0.25, and 0.5. For $\beta\!=\!0$, there is no luminosity scaling and we have $R_g({\LLstar},z)\!=\! \Rstar (z) \!\propto\! { (dN/dX(z))/f_c(z)}^{1/2}$ for a given ${\wrlim}$. In this case, the {\CIV} absorber gas radius is not physically coupled to the stellar mass and/or star formation rate (SFR). The value $\beta \!=\! 0.25$ represents an intermediate luminosity sensitivity.

\begin{figure*}[!] 
\centering
\includegraphics[width=0.97\textwidth]{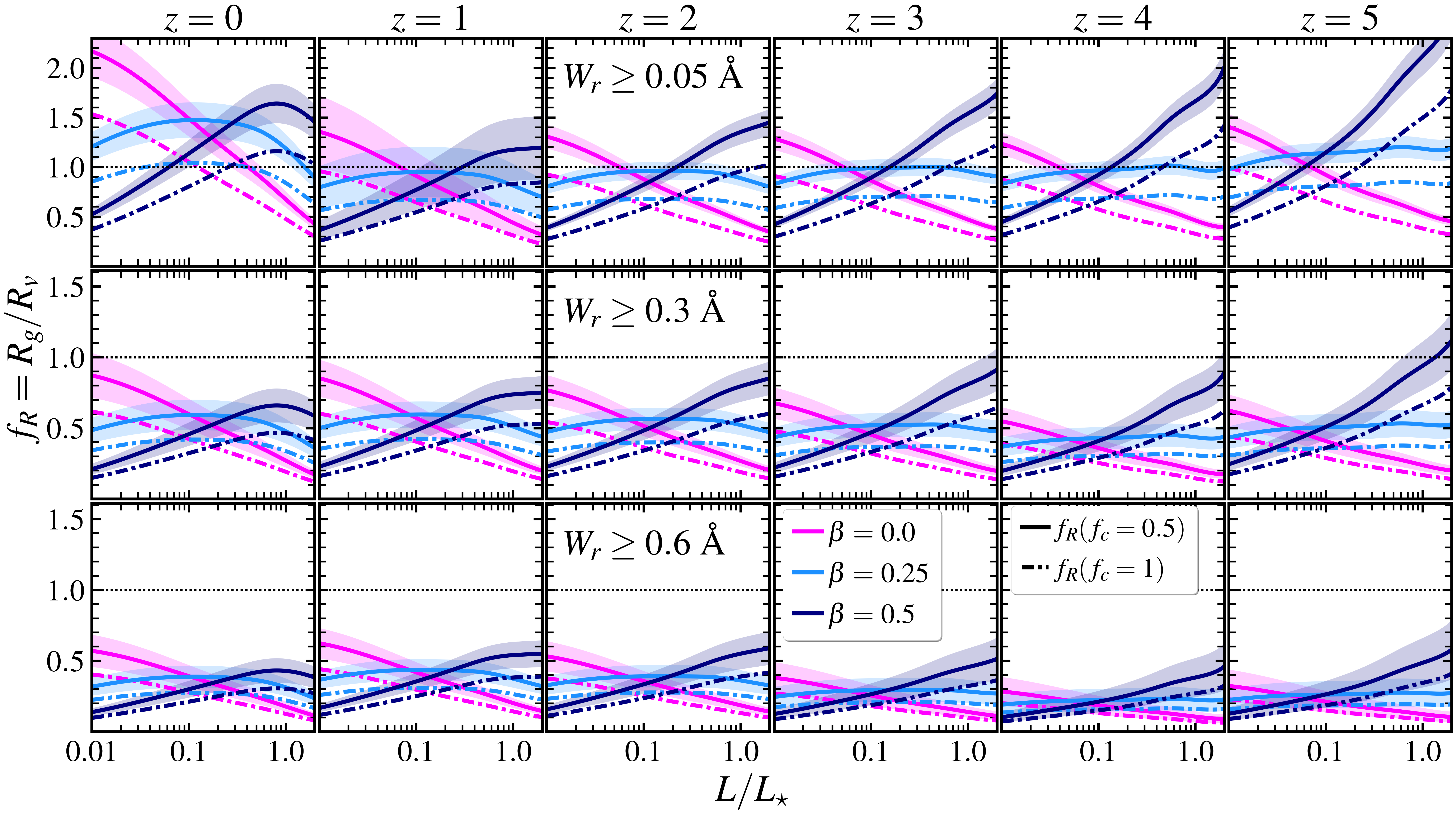}
\vspace{-5pt}
\caption{
$f_R \!=\! R_g/R_{v}$ (Eq.~\ref{eq:fR}) as a function of ${\LLstar}$ for absorbers with {\wvweak} (top panels), {\wweak} (middle panels), and {\wstrong} (bottom panels). Each column is a different redshift, from $z\!=\!0$ to $z\!=\!5$. The magenta, light blue, and dark blue curves represent $f_R$ for $\beta\!=\!0$, $0.25$, and $0.5$, respectively. For solid curves, we assume $f_c(z)\!=\!0.5$, and dash-dotted curves assume $f_c(z)\!=\!1$. ${\Lmin}\!=\!{\onepLstar}$ is adopted in every case. The horizontal dotted black line shows $f_R\!=\!1$ in each panel.}
\label{fig:fR}
\end{figure*}

In Figure~\ref{fig:fR}, we present the $\beta$ dependence of ${\fR}({\LLstar},z)$ as a function of {\LLstar} and redshift for the three {\wrlim} populations of {\CIV} absorbers. The solid curves represent $f_R({\LLstar},z)$ for $f_c(z)\!=\!0.5$, while  dashed curves are for $f_c(z)\!=\!1$. We show ${\fR}({\LLstar},z)$ for $\beta=0$ (magenta), $\beta=0.25$ (light blue), and $\beta=0.5$ (dark blue).  
As mentioned above, the \citet{Chen01} value of $\beta\!=\!0.5$ indicates a strong dependence of gas radius on galaxy luminosity, while $\beta\!=\!0.25$ indicates a weaker scaling, and $\beta\!=\!0$ means that the gas radius has no luminosity dependence.
The horizontal dotted line in all panels of Figure~\ref{fig:fR} represents ${\fR}({\LLstar},z) \!=\! 1$, where gas radius equals virial radius; ${\fR}\!>\!1$ means {\CIV} is outside {\Rv} on average, ${\fR}\!<\!1$ means {\CIV} is inside {\Rv} on average.

In Figure~\ref{fig:fR}, we see that the luminosity and redshift dependence of $f_R({\LLstar},z)$ for a given {\wrlim} is strongly determined by the choice of $\beta$. For $\beta\!=\!0.5$, the relative gas radius increases with luminosity, such that brighter (more massive) galaxies have larger {\CIV} envelopes relative to their virial radii than do fainter (less massive) galaxies. This luminosity dependence is stronger as redshift increases.
For $\beta\!=\!0$, the trend seen for $\beta=0.5$ is reversed; ${\fR}({\LLstar},z)$ increases toward lower galaxy luminosity and the sensitivity to luminosity increases as redshift decreases.
For $\beta=0.25$, the luminosity dependence of $f_R({\LLstar},z)$ is weak and changes little with redshift.

For $f_c(z)\!=\!0.5$, our analysis suggests some fraction of the weakest absorbers (with {\wvweak}) likely populate regions beyond the virial radius of galaxy halos. If $\beta\!=\!0.5$, this predominantly occurs around higher luminosity galaxies, depending on redshift; e.g. $f_R({\LLstar},z)\!\geq\!1$ for $L\!\geq\! 0.06{\Lstar}$ at $z\!=\! 0$ and $L\!\geq\! 0.2{\Lstar}$ at $z\!=\! 2$.
If $\beta\!=\!0$, super-virial gas halos are predominantly around lower luminosity galaxies; e.g. $f_R({\LLstar},z)\!\geq\!1$ for $L\!\leq\! 0.4{\Lstar}$ at $z\!=\! 0$ and $L\!\leq\!0.06{\Lstar}$ at $z\!=\!2$.

On the other hand, for $f_c(z)\!\!=\!\!0.5$, stronger absorbers typically live within the inner 50\% of the virial halos regardless of the choice of $\beta$. However, if $\beta\!=\!0.5$, our analysis would predict {\wweak} absorbers populate halos out to the outer edge of the virial radius for $L\!\geq\!{\Lstar}$ at $z\!\geq\!3$. 
For a given galaxy luminosity at redshift $z$, the relative gas radius decreases with increasing {\wrlim} (primarily because ${\dndx}(z)$ decreases with increasing ${\wrlim}$). 
The luminosity and redshift dependence of $f_R({\LLstar},z)$ for ${\wrlim}\!=\!0.3$~{\AA} and ${\wrlim}\!=\!0.6$~{\AA} absorbers are similar to that of ${\wrlim}\!=\!0.05$~{\AA} absorbers.

The above analysis has a high degree of potential for statistically constraining the spatial distribution of {\CIV} absorbers relative to the viral radius of galaxies as a function of {\wrlim}, galaxy luminosity, and redshift.  However, the resulting behavior of $f_R({\LLstar},z)$ depends strongly on the luminosity scaling of the gas radius and covering fraction.  Neither is observationally constrained as a {\it complete\/} function of luminosity or redshift. Further, the covering fraction depends on {\wrlim}, and this is not observationally established as a function of redshift. To date, we can draw only from the pioneering observational programs that examine these relationships 
\cite[][]{Chen01, Steidel10, Bordoloi14, LC14, Burchett16, Rudie19, MF21}. In this sense, until additional observations are conducted, our analysis should be viewed as an averaged approximation, assuming that such dependencies, if present \citep[see][]{Chen12}, average out.

Our estimates provide a statistical relative gas radius as a function of galaxy luminosity and redshift assuming a non-evolving luminosity scaling, $\beta$, and covering fraction, $f_c$.  For {\wstrong} absorbers, it is fairly uncontroversial that, statistically, the gas radius generally resides within the inner half of the virial halo for a wide range of luminosities over the redshift range $z\!=\!5$ to $z\!=\!0$.  For {\wweak} absorbers the regions occupied by the {\CIV} bearing gas are slightly more extended such that the $\beta$ dependence predicts the outer half of the virial radius is occupied in the low luminosity regime following Cosmic Noon ($\beta\!=\!0$) or the high luminosity regime prior to Cosmic Noon ($\beta\!=\!0.5$). 

In the case of {\wvweak} {\CIV} absorbers, we see that the luminosity scaling parameter strongly affects whether highly extended, i.e., ${\fR}({\LLstar},z) \!\geq\! 1$, absorbing structures reside around low luminosity galaxies after Cosmic Noon or around high luminosity galaxies prior to Cosmic Noon.

\subsection{Comparison with Observations}
\label{sec:compare}

\begin{figure*}[htbp]
\gridline{\fig{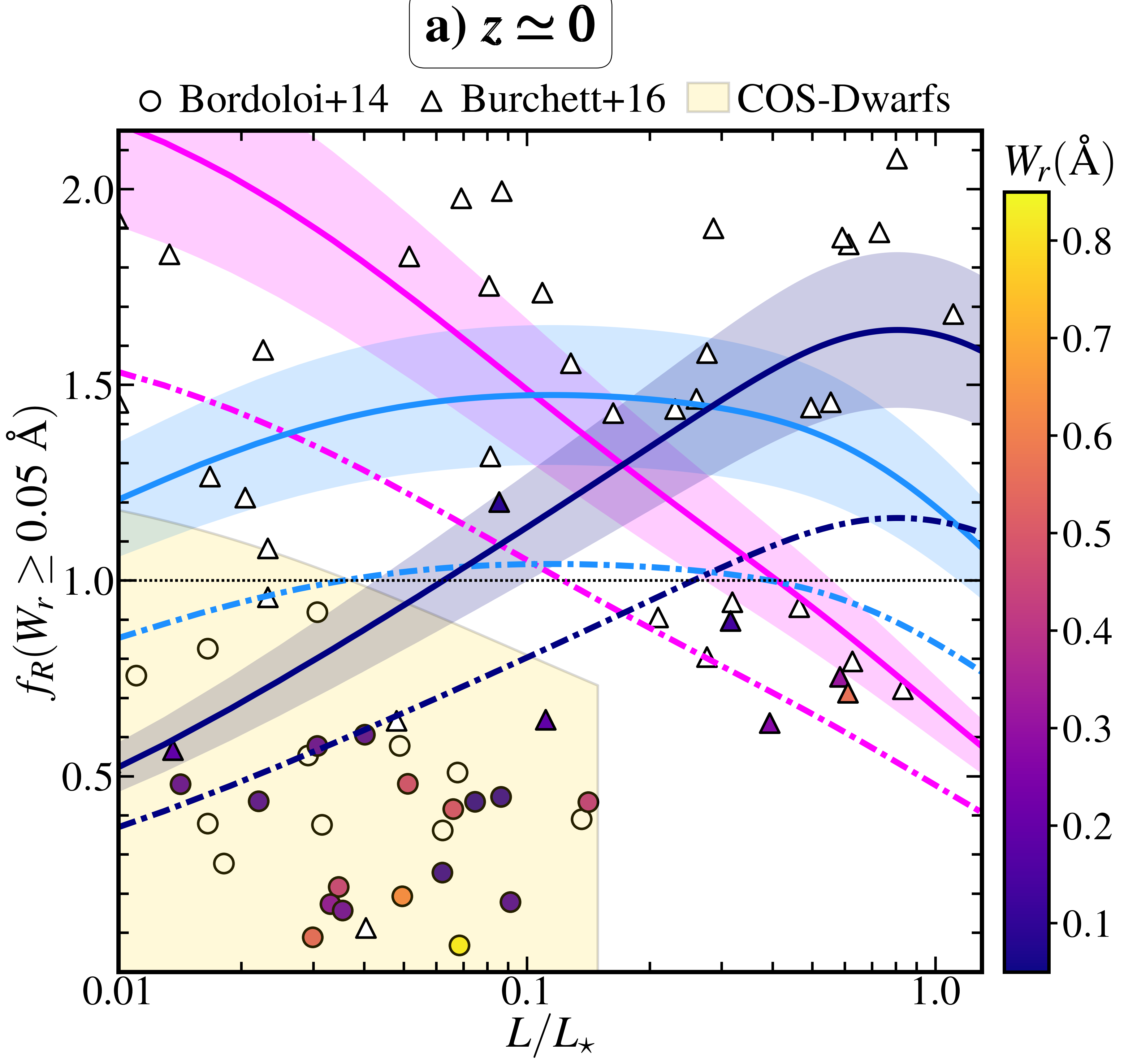}{0.34\textwidth}{}
 \fig{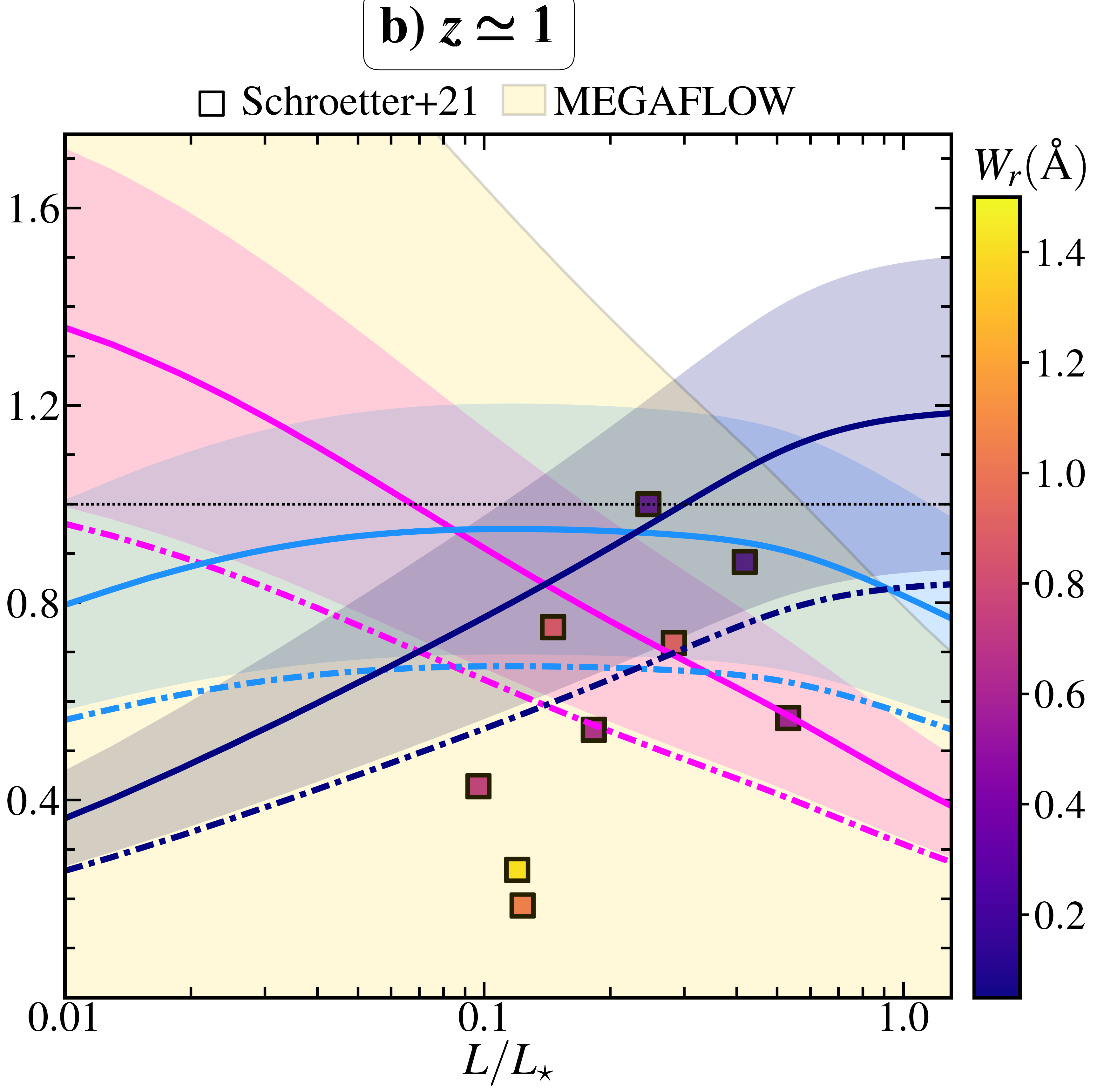}{0.32\textwidth}{}
 \fig{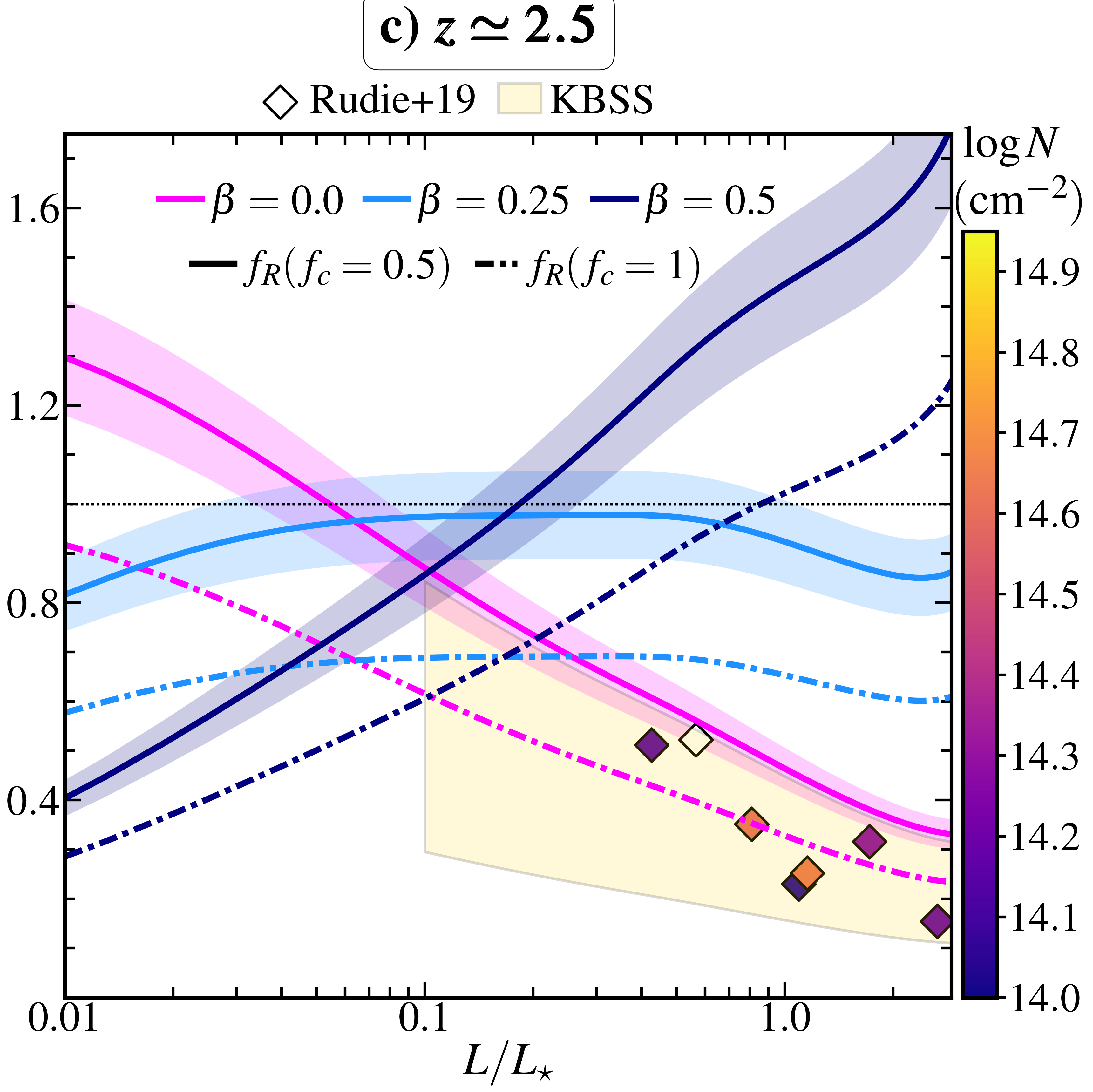}{0.32\textwidth}{}
        } \vspace{-25pt}
\caption{$f_R({\LLstar},z)$ for absorbers with {\wvweak} as a function of galaxy luminosity, at (a) $z\!\simeq\!0$, (b) $z\!\simeq\!1$, and (c) $z\!\simeq\!2.5$. As in Figure~\ref{fig:fR}, the magenta, light blue, and dark blue curves in each panel show the $f_R$ estimated from eq.~\ref{eq:fR} with $\beta \!=\! 0$, $0.25$, and $0.5$, respectively. The solid and dash-dotted curves represent covering fractions of $50\%$ and $100\%$, respectively. (a) The circles are observed $f_R$ values for COS-Dwarfs galaxies \citep{Bordoloi14}, while the triangles are from the sample of \citet{Burchett16}.  All the detections are color-coded by {\CIV} absorber {\EWr}, while the non-detections are shown by open circles and triangles.
(b) The squares are observed $f_R$ values from \citet{MF21}, color-coded by absorber {\EWr}. (c) The diamonds are observed $f_R$ values from \citet{Rudie19}, color-coded by absorber column density, $\log N~(\mathrm{cm}^{-2})$.
The yellow shaded areas represent the effective search space of (a) the COS-Dwarfs ($\!D\leq\!150$~kpc and ${\LLstar}\!\leq\!0.15$), (b) MEGAFLOW ($D\!\leq\!250$~kpc), and (c) KBSS ($D\!\leq\!100$~kpc and ${\LLstar}\!\geq\! 0.1$) surveys, respectively. All these studies had a detection threshold of ${\wrlim} \!\simeq\! 0.05$~{\AA}.} 
\label{fig:fR_comp}
\end{figure*}

Hereafter, we refer to ${\fR}({\LLstar},z)$ as {\fR}, such that the luminosity and redshift are implied in the ratio ${\Rg}/{\Rv}$.
In Figure~\ref{fig:fR_comp}, we compare the available observational data to our estimates of $f_R$ for three different $\beta$ values ($\beta\!=\!0$ is magenta, $\beta\!=\!0.25$ is light blue, and $\beta\!=\!0.5$ is dark blue) and two covering fractions ($f_c(z)\!=\!1$ are dash-dotted and $f_c(z)\!=\!0.5$ are solid).

In Figure~\ref{fig:fR_comp}(a), we plot the absorber-galaxy pairs of \citet[][circles]{Bordoloi14} and \citet[][triangles]{Burchett16} at $z\!\simeq\!0$. Since both studies adopted $R_{200}$ for the virial radius, we applied a scale factor of $\sim$1.2 to enable direct comparison with {\Rv} (see Appendix~\ref{app:obsdata}). 
The COS-Dwarfs survey \citep{Bordoloi14} selected for $L \!\leq\! 0.14{\Lstar}$ galaxies probed within impact parameters of $D\!\leq\!150$~kpc (their search space is shaded light yellow). The blind survey of \citet{Burchett16} searched for galaxies associated with {\wvweak} {\CIV} absorbers with no {\it a priori\/} selection for galaxy luminosity or impact parameter. Note that the measured $W_r$ is color coded over the range shown on the color bar.

Given the number of non-detections (open data points) below the {\fR} curves, it is clear that $f_c\!=\!1$ does not apply. 
For $f_c\!=\!0.5$, taking a census of the data, we find that for $\beta\!=\!0$ the observed covering fraction within ${\fR}$, i.e., under the magenta curve, is $f_c\!=\!N_{det}/N_{tot} \!=\! 0.46$. Similarly for $\beta\!=\!0.25$, the light blue curve, we count $f_c\!=\!0.48$, and for $\beta\!=\!0.5$, the dark blue curve, we count $f_c\!=\!0.56$.
However, dwarf galaxies with $L\!\leq\!{\tenpLstar}$ are found to typically contain {\wvweak} {\CIV} in the inner half of the virial halo (${\fR}\!\leq\!0.5$) and only one such galaxy has ${\fR}\!>\!1$ as is predicted by a $\beta\!=\!0$ scenario.
Thus, the data at $z\!\simeq\!0$ are consistent with $\beta\!=\!0.25$ and $\beta\!=\!0.5$ for $f_c\!=\!0.5$, but not with $\beta\!=\!0$ at low luminosities.
These covering fractions are in reasonable agreement with those found by \citet{Bordoloi14}, \citet{LC14}, \citet{Burchett16}, and \citet{Manuwal21} of $f_c\!=\!0.5$, $0.45$, $0.33$, and $0.43$ respectively, within the virial radius.  \citet{Chen01} found $\beta\!=\!0.5$ for a sample of galaxies with $\langle z \rangle \!=\! 0.4$, but this was for $B$-band luminosity (we adopted the UV-band) and for a {\CIV} absorption sensitivity of $W_r \!\geq\! 0.2$~{\AA}. \citet{nikki13b} found the {\MgII} $\beta$ to vary significantly with luminosity band and {\wrlim}. 
At $z\!\simeq\!0$, the most leverage that could be applied to constrain our models would occur by increasing the number of detections at $f_R \!\sim\! 2$ for both $L \!<\! 0.1{\Lstar}$ and $L \!\simeq\! {\Lstar}$.

In Figure~\ref{fig:fR_comp}(b), we compare to the observations of \citet{MF21} covering $1.0 \!\leq\! z \!\leq\! 1.5 $. These data correspond to the {\CIV} absorbers in their study for which the authors identified a single galaxy ``counterpart.''
The methods by which we computed the virial radii and UV band luminosities of these galaxies are presented in Appendix~\ref{app:obsdata}. The \citet{MF21} search space was 250 pkpc (proper kpc) projected from the background quasar at the redshift of the absorption and their minimum luminosities extend below 0.01{\Lstar} (their search space is shaded light yellow). Further, the absorption was known {\it a priori}, so we are plotting detections only. However, from their survey data, they presented an impact parameter dependent model of the covering fraction, which we estimate to have an impact parameter weighted average of $f_c\!=\!0.25$ within {\Rv}.

Of the nine galaxies appearing on Figure~\ref{fig:fR_comp}(b), two are inconsistent with $\beta \!=\! 0$ for $f_c\!=\!0.5$.  However, these two galaxies are consistent with $\beta \!=\! 0.25$ and 0.5, especially if the covering fraction is lower than $f_c\!=\!0.5$. At $1.0 \!\leq\! z \!\leq\! 1.5$, the most leverage that could be applied to constrain our models would occur by increasing the number of detections to $f_R \!>\! 1$ for $L \!<\! {\tenpLstar}$.  However, continued examination of $L \simeq {\Lstar}$ galaxies out to $f_R \!>\! 1.2$ would also apply leverage.  As our formalism associates a single absorber to an individual galaxy, we present only the {\it single} galaxy counterparts to {\CIV} absorbers from \citet{MF21}, which constituted just $\approx$17\% of their detected {\CIV} sample. In contrast, $\approx$30\% of their absorbers had multiple galaxy counterparts within 250 kpc, indicating the role that clustering or grouping of galaxies may play in {\CIV} absorption. We revisit this observational fact and its implications in detail below.

Finally, in Figure~\ref{fig:fR_comp}(c), we compare to the seven galaxies observed by \citet{Rudie19}, covering $2.1 \!\leq\! z \!\leq\! 2.7$. The impact parameter range for this targeted sample of galaxies was 35--100 pkpc projected from the background quasar at the redshift of the absorption (their search space is shaded light yellow). The methods by which we computed the virial radii and UV band luminosities of these galaxies are presented in Appendix~\ref{app:obsdata}. The measured absorption strengths are not {\EWr}, but are the total {\CIV} column densities (sum of Voigt profile components), color coded over the range shown on the color bar.  For reference, a single-component {\CIV} absorber with $\log N/{\rm cm}^{-2} \!=\! 14.1$ would have $W_r \!=\! 0.13$, $0.2$, and $0.3$~{\AA} for $b\!=\!8$, $15$, and $30$~{\kms}, respectively.

Weighting by column density, we estimate that the \citet{Rudie19} sample has a mean covering fraction of $f_c \!\simeq\! 0.75$ for $\approx${\Lstar} galaxies at $2.1 \!\leq\! z \!\leq\! 2.7$ (see their Figure~5). Though the galaxies appear nearly consistent with $\beta\!=\!0$ and $f_c\!=\!0.5$, this is an artifact of the galaxies being selected within a luminosity independent, fixed impact parameter range (100 pkpc). Unfortunately, the data available at the Cosmic Noon era shown on Figure~\ref{fig:fR_comp}(c) fill too limited of a parameter space to discriminate between the free-parameters $\beta$ and $f_c$ in our models.  At these redshifts, the derived UV luminosities of the galaxies are sensitive to $k$-corrections, which can range from 0.2 for an Im galaxy, 1.2 for an Scd galaxy, and 2.2 for an Sbc galaxy \citep[e.g.,][]{nikki13a}. As described in Appendix~\ref{app:obsdata}, we adopted the average of Im and Scd galaxies, appropriate for star forming galaxies at these redshifts.

To improve the observational constraints on $\beta$ and $f_c$, much larger samples of direct detections to small equivalent width detection thresholds ($W_r \!\simeq\! 0.05$~{\AA}) will be necessary. In summary, at $z\!\sim\!0$, observations are needed to $f_R \!\sim\! 2$ for $L \!<\! {\tenpLstar}$ and $L \!\simeq\! {\Lstar}$.  At $1.0 \!\leq\! z \!\leq\! 1.5 $, observations to $f_R \!>\! 1$ for $L\!<\!{\tenpLstar}$ would be useful. At $2.1 \!\leq\! z \!\leq\! 2.7$, the data remain highly incomplete; at Cosmic Noon it will remain a challenge to constrain the $\beta$ luminosity dependence unless $\beta \!\geq\! 0.5$ and deep surveys down to ${\LLstar} \!\leq\! 0.1$ out to ${\fR} \!=\! 1.2$ can be conducted.  These data are critical as they would provide direct information on where {\CIV} lives around galaxies of different luminosities as a function of redshift. Such data would provide essential observational constraints on cosmological simulations that theoretically model the baryon cycle of galaxies.

\subsection{Uncertainties, Assumptions, and Caveats}
\label{sec:assumptions}

Our primary assumption in calculating the {\CIV} gas radius is that each absorber is found around a single galaxy, so that the abundance of absorbers is the same as the abundance of galaxies above a minimum luminosity and of DM halos above a minimum mass. However, several pieces of observational evidence, which we discuss in Section~\ref{sec:modelcontext}, would indicate that this may not be an accurate picture of galaxy-absorber association. 

The Holmberg luminosity scaling, $\beta$, used to estimate gas radius (Eq.~\ref{eq:rstar}) is quite uncertain. To date, only \citet{Chen01} has constrained $\beta=0.5$ for $B$-band luminosity at $z\!<\!1$ for {\CIV} with $W_r \geq 0.2$~{\AA}. Due to the lack of sufficient constraints, we have set $\beta$ to be a fixed quantity for all {\wrlim} at all {\Lstar} and all $z$.  However, for {\MgII}, \citet{nikki13b} established that $\beta$ varies with $z$, {\wrlim}, galaxy color, and luminosity band.  Furthermore, \citet{cwc13} reported different halo mass $\beta$ scalings as {\MgII} {\wrlim} is varied. Lastly, {\CIV} absorption is known to be more extended for AGN-host galaxies than other galaxy types \citep[e.g.][]{Prochaska14}, which would imply that these galaxies may have a different Holmberg scaling. 

The covering fraction $f_c$ is an additional uncertainty. In Eq.~\ref{eq:rstar}, we have assumed a ``mean'' covering fraction for gas halos at a given redshift. In fact, the {\CIV} covering fraction is observed to have a profile that declines with increasing distance from the central galaxy \citep[e.g.,][]{Bordoloi14,Burchett16, MF21}. Furthermore, at fixed impact parameter, $f_c$ is observed to  decrease with increasing {\wrlim} \citep{Bordoloi14,Rudie19}. The few existing measurements are also limited by small sample sizes, especially at $z\!>\!2$ \citep[e.g.,][]{Rudie19}. The covering fraction is also found to be dependent on star formation activity and stellar/halo mass \citep{Bordoloi14,Burchett16}, implying that $f_c$ is also potentially a function of luminosity or halo mass. In addition, the covering fraction of gas around quasar-host galaxies is known to be higher than around other types of galaxies \citep[e.g.][]{Prochaska14,Landoni16}. For example, \citet{Prochaska14} found ${\EWr}\!>\!0.2$~{\AA} {\CIV} to persist beyond $\sim$500~kpc from $M_h \!\gtrsim\! 10^{12.5}$~{\Msun} quasar-host galaxies at $z\!\sim\!2$. However, given that such galaxies are very massive ($L\!>\!{\Lstar}$), they represent a much rarer population of galaxies than the overall population of galaxies at $L\!\sim\!0.01$--2{\Lstar} that we study here. Finally, \citet{Burchett16} also reported a dependence of covering fraction on galaxy environment. Given the lack of robust constrains on $f_c(z)$, we assumed a constant $f_c$ for all {\LLstar} and for all {\wrlim}, without regard to impact parameter.

The minimum galaxy luminosity, {\Lmin}, as applied in both Eq.~\ref{eq:rstar} and our DM halo abundance matching is also not well-constrained. The low-$z$ studies of \citet{Bordoloi14} and \citet{Burchett16} suggest that {\wvweak} {\CIV} is found around galaxies as faint as $\sim\!{\onepLstar}$, but this limit is found to be about $\sim\!0.4{\Lstar}$ at $z\!\sim\!2.5$ \citep[][see Figure~\ref{fig:fR_comp}]{Rudie19}. It remains unclear whether {\Lmin} decreases at higher redshift or if an apparent decrease is simply due to the difficulty in identifying fainter galaxies with {\CIV} absorbers at higher redshifts. We assumed ${\Lmin}\!=\!{\onepLstar}$ for all redshifts, based on {\CIV} detections around these lower luminosity galaxies at low redshifts. Assuming ${\Lmin}\!=\!{\tenpLstar}$ instead yields $\approx\!25$--$40$\% smaller {\fR} values at $z\!=\!2$--5, but the redshift-evolution is qualitatively the same. Hence, the magnitude of gas radius values is sensitive to {\Lmin}, but the evolution is not.

Central to our model is the assumption that a single luminous galaxy resides in each DM halo.  To equate luminous galaxies to DM halos, we adopted abundance matching, which provides a one-to-one mapping between galaxy luminosity and DM halo mass for a given adopted luminosity function and halo mass function (see Appendix~\ref{app:abundance}). While we do incorporate appropriate scatter to account for the vast array of baryonic processes involved in galaxy formation in halos, there are several additional sources of systematic and statistical uncertainty in abundance matching that merit detailed discussion \citep[see][]{Behroozi10,WT18}.

The model also does not take into account any halo mass dependence of {\wrlim}, such that all populations of absorbers, regardless of absorption strength, are assumed to have the same abundance as halos above a minimum mass. This neglects any possible underlying relationship between halo (or galaxy) mass and absorption strength, for example if strong {\CIV} absorbers were found only around high-mass galaxies.

The modelled behavior of the of gas radius relative to the virial radius depends on the adopted definition of the virial radius. While many authors employ a fixed virial overdensity of $\Delta_v\!=\!200$, we use the spherical collapse approximation of \citet{BN98}, where $\Delta_v$ ranges from $\approx$370 (at $z\!=\!0$) to $\approx$178 (at $z\!>\!4$) for our cosmology \citep{WMAP09}. Using the most recently published set of cosmological parameters \citep{Planck20} would change the resulting {\Rv} by up to $\sim$4\%. 
Some authors have used alternative definitions for the extent of influence of a DM halo. \citet{More15} advocated the use of a ``splashback radius,'' which corresponds to the apocenter of the orbits of the most recently accreted matter in a halo, as a more physically motivated halo boundary than the virial radius. \citet{Shull14} instead considered the redshift at which half the mass in a halo had collapsed/virialized, which yields a $\approx$40--50\% smaller virial radius. Finally, \citet{Diemer13} demonstrated that at $z\!<\!1$, a large fraction of the growth of halo mass and virial radius is due to pseudo-evolution, i.e. the evolution of a reference density (such as the critical density), rather than the actual physical accretion of matter.

Finally, we note that our model is based on many average properties of galaxies and DM halos as a function of redshift.  We have taken no account of galaxy environment nor the rich diversity of individual galaxy formation and evolution histories (including the star formation history which would dictate the strengths and rates of outflows; see below) that certainly have a correspondence with how {\CIV} absorbing gas is distributed on a case by case basis.


\section{Discussion} 
\label{sec:discuss}

We investigated the statistical connection between galaxy halos and their enriched gas envelopes across $\sim$12.5~Gyr of cosmic time, based on observed statistics of {\CIV} absorption selected systems and galaxies. While we assume a simplified model in which each individual {\CIV} absorber is associated with a single galaxy DM halo, we recognize that this scenario represents only one of many possible scenarios for how metal-enriched absorbers observed in quasar spectra are related to galaxies and DM halos.  Our model does serve, however, to provide a working framework on which to examine this relationship as more data become available in the future.  In this section, we examine our model in the context of the baryon cycle governing galaxy evolution and then discuss several alternative scenarios and how they compare.

\subsection{Our Model in Context}
\label{sec:modelcontext}

The assumption that each absorber is found around a different galaxy, hence of equal abundance of these populations, is a stringent criterion of our model. {\CIV} absorbers found in quasar spectra are sometimes clustered close to each other in redshift space, indicating they could arise in the CGM of the same galaxy \citep[e.g.,][]{Peroux04,Cooksey13,Burchett15,H20,MF21}. Significant clustering of absorbers in physical space could result in degeneracies in the absorber abundance relative to galaxy abundance. However, the clustering of systems does depend on the velocity window across which a single system is defined, and this varies among different studies. For context, we defined a single absorber to be within a velocity window of $\Delta v \!=\! 500$~{\kms} \citepalias{H20}. In our current sample, $\sim$1\% and $\sim$5\% of the absorbers are found within 1000 and 3000 {\kms}, respectively, of another absorber along the same line-of-sight. 
While this could imply possible physical connection between those absorbers, such as associated via a large-scale structure such as a cluster or group, the large velocities are consistent with expectations that a negligible number of our absorbers overlap in a single DM halo.

Transverse absorber studies have found a non-negligible fraction of metal absorbers beyond the virial radii of galaxies \citep[e.g.,][]{Prochaska11, Turner14, Mathes14, LC14, Huang21} and/or gravitationally unbound from DM halos due to their apparent velocities \citep[e.g.,][]{Tumlinson11, Bordoloi14, Rudie19}. On the other hand, \citet{MF21} found that a majority of their {\CIV} absorbers and $\sim$73\% of the absorbers with only {\CIV} (but no {\MgII}) detected did not have an associated galaxy counterpart within 250 kpc down to a sensitivity of ${\LLstar} \!\simeq\! 0.01$ at the redshift of the absorbers ($1 \!\leq\! z \!\leq\! 1.5$). If a substantial fraction of {\CIV} absorbers live in the IGM, then the assumption of each absorber being associated with a galaxy DM halo would break down. However, this phenomenon would conform with our inference that some weak absorbers reside beyond the virial radius of DM halos. Alternatively, the \citet{MF21} result could be explained by absorption from the CGM of lower mass or quiescent galaxies below their luminosity detection threshold.

The model also does not account for systematic relationships between {\CIV} absorber strength and stellar activity, such as specific star formation rate.  For example, \citet{Borthakur13} found a higher incidence rate of {\CIV} absorption around star-bursting galaxies than around their control sample.  Incorporating the sample of \citet{Borthakur13}, \citet{Bordoloi14} showed that higher {\EWr} {\CIV} absorbers are preferentially found around galaxies with higher specific star formation rates.

However, it is likely that the biggest challenge to our adopted model arises from the lack of accounting for overlap between virial halos. We treat DM halos as individual, isolated objects with their separate abundances and cross-sections. If there is significant overlap between different halos, the abundances would be different from what we calculated for isolated low-mass and high-mass halos. Several studies of the spatial distribution of intergalactic metals using semi-analytical outflow models have concluded that strong clustering of enriching sources typically leads to significant overlap between enriched regions \citep[e.g.,][]{Scanna05, Scanna06, Samui08}. 
Furthermore, observations indicate that {\CIV} absorption is affected by the large-scale environment of galaxies \citep[e.g.,][]{Adelberger03,Adelberger05,Burchett16}.
We further discuss this issue below.

\subsection{Baryon Cycle and the CGM} \label{sec:cgmdiscuss}

From a phenomenological standpoint, as star formation increases, more metals are produced, and more metal-enriched gas is therefore driven out of galactic centers via outflows. Therefore, an {\it a priori} expectation for metal-line absorbers is that their cosmic incidence should increase in tandem with global star formation activity of the universe. 
Indeed, strong {\MgII} absorber incidence is found to increase from $z\!\sim\!5$ to a peak around $z\!\sim\!2$ and fall afterwards \citep[e.g.,][]{MS12}, in correspondence with the global cosmic star formation rate density \citep[e.g.,][]{MD14} and the increased incidence of outflows \citep[e.g.,][]{Rupke18} at Cosmic Noon. Similarly, the redshift-evolution of {\wweak} and {\wstrong} {\CIV} absorbers is qualitatively consistent with the evolution of cosmic star formation activity at $z\!>\!2$, implying a possible causal connection between strong {\CIV} and global star formation rate at early epochs, as was argued by \citet{ZFChen16}. 

However, as seen in Figure~\ref{fig:dndx_halo+abs}, the cosmic incidence of {\wvweak} {\CIV} absorbers does not evolve in this fashion. Instead, it rises rapidly in the last 8 Gyr {\it following} Cosmic Noon, having been effectively constant across ${1.5 \!\leq\! z \!\leq\! 2.5}$. In the context of our models, the relative extent ($f_R = R_g/R_v$) of this population of absorbers around galaxies may even be at its global minimum just following the epoch of Cosmic Noon (see Figures~\ref{fig:lstar} and \ref{fig:10lstar}).  In contrast, the incidences of both stronger {\CIV} with {\wweak} and {\wstrong} monotonically, yet slowly, increase across Cosmic Noon.

The most rapid evolution is for {\wvweak} {\CIV} at $z\!<\!1$. This late-time rapid evolution is difficult to explain physically, as neither the metal density relative to the cosmic mean nor the expected amount of metals based on star formation has increased rapidly in the last 8~Gyr \citep[e.g.,][]{PH20}. Furthermore, the ionizing background at the {\HeII} ionization edge (which is similar to the ionization energy of {\CIV}) has softened by $z\!\sim\!1$, meaning {\CIV} photoionization conditions have not been particularly favorable in these last 8 Gyr \citep[e.g.,][]{HM12}. \citet{Borthakur13} suggested a model for $z \!\leq\! 0.2$ in which {\CIV} would arise in pre-existing clumpy clouds or filaments in the CGM that are being shock-ionized and accelerated by the ram pressure of winds from nearby starburst galaxies. They argue that the ensemble of such clouds must have a  covering fraction of $\sim$80\%.  It remains to be worked out whether such a scenario can be made consistent with the rapid evolution in very weak {\CIV} at $z\!<\!1$.

However, the rapid rise in {\dndx} (and subsequently rapid rise in {\fR}) of {\wvweak} {\CIV} at $z\!<\!1$ could be an effect of increasing {\it collisional} ionization of {\CIV} instead of photoionization. Cosmological hydrodynamical simulations have predicted the increasing prevalence of collisional ionization of metals with decreasing redshift, as halos that grow to higher masses ($M_h \!\gtrsim\! 10^{12}~{\Msun}$) have higher virial temperatures ($T_v \!\gtrsim\! 10^{5}$~K) able to collisionally ionize metals to higher ionization states \citep[e.g.,][]{OD06,Oppenheimer16,Oppenheimer18}. In the EAGLE simulations, at $z\!\leq\!0.25$, the {\OVI} ionization fraction peaks at $\sim$1--2{\Rv} from $L\!\sim\!{\Lstar}$ halos owing to the optimal temperature for collisional ionization to {\OVI} \citep{Oppenheimer16}. The lower virial temperature of sub-{\Lstar} halos results in a lower {\OVI} fraction, while higher virial temperatures of group-sized ($L\!\gtrsim\!2{\Lstar}$) halos are able to further ionize {\OVI} to higher states such as {\OVII} and also result in a lower {\OVI} fraction. \citet{OD06} found large fractions of collisionally ionized {\CIV} absorbers in their simulations due to favorable virial temperatures of $T\!\simeq\!10^5$~K, a trend accentuated with decreasing redshift (especially at $z\!<\!2.5$).

Collisional ionization induced by hot gas at or outside the virial radius of massive galaxies could explain how {\wvweak} {\CIV} incidence increases in the last $\sim$8 Gyr and how the relative gas radius grows significantly from $z\!=\!1$ to $z\!=\!0$. 
Additionally, the fact that the {\wvweak} {\fR} peaks at $L\!\approx\!{\Lstar}$ and drops for higher luminosities at $z\!=\!0$ (see top left panel of Figure~\ref{fig:fR}, $\beta\!=\!0.5$ curve) might be explained by the \citet{Oppenheimer16} conclusion that more massive halos would collisionally ionize metals to higher states (although the collisional ionization temperature of {\CIV} is $\sim\!0.5$~dex lower than that of {\OVI}). On the other hand, lower ionization metals are primarily photoionized and subject to very little collisional ionization, even at lower redshifts \citep[e.g.,][]{Ford13}.
Empirically, \citet{BenR18} showed that photoionization cannot produce detectable {\OVI} absorption at $z\!\leq\!0.2$ due to the decreasing strength of the UVB at low redshifts, instead requiring collisional ionization to generate such high-ionization absorbers at late epochs. For {\CIV}, there is a similar expectation of collisional ionization dominating at low $z$ (however see \citet{Manuwal21}), while lower ionization metals such as {\MgII} and {\FeII} remain mostly photoionized. 
The increasing prevalence of collisional ionization of higher ionization metals but not lower ionization metals, in tandem with the decreasing UVB at later times, could potentially explain why the {\CIV} {\dndx} rises rapidly at $z\!<\!1$ but {\MgII} {\dndx} falls at those redshifts \citep{MS12,H20}.

The evolution of the statistical gas radii provides a more detailed look into how {\CIV} absorbing structures grow relative to galaxies (see Figures~\ref{fig:lstar} and \ref{fig:10lstar}). For {\Lstar} and $0.1{\Lstar}$ galaxies, we find that the relative gas radius for {\wstrong} {\CIV} absorbers slightly grows from $z\!\sim\!3$ to a peak at $z\!\sim$1.5--2, and then declines afterwards. This suggests a causal physical connection between {\wstrong} {\CIV} around galaxies and star formation processes, similar to, yet more mild, than what is found for strong {\MgII} absorbers \citep{MS12,Chen17}.

The intermediate {\wweak} {\CIV} population exhibits relatively little evolution. The relative gas radii evolves similarly if somewhat less dramatically as that of the {\wstrong} {\CIV} absorbers.  Both evolve differently from the {\wvweak} population. In contrast, the relative gas radii of {\wvweak} {\CIV} absorbers appear to evolve independent of the star formation peak across the Cosmic Noon epoch.

These differences in evolution may be suggestive of different origin scenarios for {\CIV} absorbers of different absorption strength. 
In particular, star formation-driven outflows could be a major pathway to populating the inner regions of DM halos with {\wstrong} {\CIV} absorbers. And in fact, there is some observational evidence connecting strong {\CIV} absorbers to galactic winds \citep{Fox07, Steidel10, nikki20}.

Since the gas radius of {\wvweak} {\CIV} absorbers relative to the virial radius appears to evolve independently of (if not opposite to) the global cosmic star formation history, the weakest absorbers would seem less likely to be tracing intrinsic feedback processes like outflows. Though this statement is counter to the interpretations of \citet{Borthakur13} at $z\! \simeq \!0$, a possible alternative scenario is that the gas halos of {\wvweak} {\CIV} absorbers may grow via extrinsic processes, such as changing environmental conditions in the IGM in the vicinity of DM halos that favor the C$^{+3}$ ionization state and small column densities (for example, local overdensity evolution of the outer halo/IGM interface or enrichment by star-forming sub-halos). Alternatively, the {\wvweak} {\CIV} gas radius may trace distant-past star formation activity that is eventually manifested (post Cosmic Noon) as an evolving increase in weak {\CIV} absorption in the outer extremes of the virial halo.

Such scenarios seem reasonable given the remarkably similar evolution of the gas radius of {\wvweak} {\CIV} absorbers and the virial radius until $z\!\sim\!1$ for {\Lstar} galaxies (see Figure~\ref{fig:lstar}(a)). 
Global statistics of absorbers, galaxies, and DM halos would suggest that, depending on the covering fraction, the gas radius for this population of absorbers resides near or beyond the virial radius of {\Lstar} galaxies for the majority of cosmic time.
For {\tenpLstar} galaxies, we find similar results with the exception that the gas radius is not as extended relative to the virial radius (see Figure~\ref{fig:10lstar}(a)). 
However, the rapid growth of the radius of {\wvweak} absorbers relative to the virial radii of both {\tenpLstar} and  {\Lstar} galaxies over the last 8~Gyr suggests a shift in the dominant mechanism for the creation and sustenance of a weak {\CIV} absorber population following Cosmic Noon.

Using a similar formalism to ours, \citet{Schaye07} calculated the characteristic gas radius of {\CIV} clouds with super-solar metallicity at $z\!\sim\!2.3$. Assuming ${\Lmin}\!=\!{\tenpLstar}$, they estimated ${\Rstar} \!\approx\!80$ kpc, which is remarkably close to the value of {\Rstar} we find for {\wstrong} absorbers (see Figure~\ref{fig:lstar}(c)). This correspondence of absorbing halo sizes indicates that the strongest {\CIV} absorbers are metal-rich and therefore likely inhabit denser regions of the CGM. On the contrary, {\Rstar} for {\wvweak} is about a factor of three larger at this redshift, suggesting weak {\CIV} absorbers live in less enriched and therefore more tenuous regions.

In a recent comparative analysis of a diverse suite of cosmological simulations, \citet{Fielding20} found that different physical processes shape the inner and outer regions of the CGM of $\sim${\Lstar} galaxies at low redshifts. Intrinsic galactic feedback plays the dominant role in shaping the inner CGM, while extrinsic processes such as cosmological accretion and interaction with satellites dominate the makeup of the outer CGM. \citet{Li21} also established the distinction between an outflow-dominated inner CGM and accretion-dominated outer CGM in dwarf galaxies in the FIRE simulations.  
Our result that the characteristic extent of strong {\CIV} is statistically confined to the inner CGM and very weak {\CIV} extends to the outer halo and beyond, is consistent with these theoretical works that suggest stronger absorbers are primarily associated with intrinsic feedback while weaker absorbers mostly trace extrinsic IGM accretion/satellite interaction processes.

In a comprehensive study of absorbing gas dynamics, origins, and fates at $z\!=\!0.25$, \citet{Ford14} quantified the fraction of each absorbing species originating as outflow, newly accreted, recycled, or ambient gas from the IGM, as a function of both halo mass and projected galactocentric distance (impact parameter). The {\CIV} absorbing gas in the inner halos is dominated by ``recycled accretion'' (accreting gas that was ejected in a wind at least once in the past), whereas the outer halos are dominated by ancient outflows (non-accreting gas ejected into the CGM by a galactic wind that occurred more than $1$~Gyr in the past). At $100$~kpc, in the CGM of both $10^{11}$~{\Msun} and $10^{12}$~{\Msun} halos, more than 50\% of the weak {\CIV} absorption originated from ancient outflows.

Interestingly, \citet{Ford14} also found that ``young outflows'' (non-accreting gas younger than 1 Gyr) significantly contribute to strong {\MgII} absorption (especially in the inner halo where strong {\CIV} absorbing gas is predominantly recycled accretion).  These simulations naturally explain why the strong {\MgII} absorber cosmic incidence closely traces the global cosmic star formation \citep{MS12} and why the strong {\CIV} absorber incidence does not \citepalias[\citealt{Cooksey13};][also see Figure~\ref{fig:dndx_halo+abs} of this work]{H20}. That the gas radius of strong {\CIV} is confined to the inner halo and that, relative to the virial radius, it shows only a mild growth during the Cosmic Noon epoch, would be consistent with {\CIV} tracing mostly recycled accretion, as opposed to young outflows. 
\citet{Oppenheimer20} found that the {\CIV} covering fraction shows a delay of $\sim$0.5--2.5~Gyr to respond to gas ejection into the CGM, in contrast to the much quicker response from covering fraction of low-ionization phases.

A general picture appears in which higher column density {\CIV} bearing gas is predominantly recycling back onto the galaxy after $\sim$1 to a few billion years of transiting through the inner halo, while lower-column density {\CIV} bearing gas resides in the outer halo primarily as non-accreting ancient outflow material. One aspect of this picture is that star-forming (satellite) sub-halos are not taken into account in the work of \citet{Ford14}. There is clearly the possibility that {\CIV} bearing gas can be seeded into the outer regions of more massive central halos from nearby lower-mass satellites.

Taking a  broader view, simulations find that CGM and IGM gas is substantially mixed so that a given background quasar line of sight may intercept absorbing gas with multiple origins and fates \citep[e.g.,][]{Ho20,Li21}. As such, our attempts to infer that absorbers are associated with a single origin or fate are probably naive due to the true complexity of the baryon cycle. There may be very little causal correlation between the origin and future evolution of absorbing gas traced by a single ion \citep[e.g.,][]{Hafen19, Hafen20}. Yet, it is of continued interest to attempt to understand the origin and fate of the {\it bulk\/} of the selected metal-line absorbing gas in DM halos that host galaxies.

\subsection{Baryon Cycle and Galaxy Evolution} \label{sec:galdiscuss}

Models that track the chemical enrichment history of the universe generally agree that low-mass galaxies are the principal polluters of the IGM, especially at early times. \citet{Booth12} determined that observational constraints on the median {\CIV} optical depth can be satisfied only if the IGM was primarily enriched as early as $z\!=\!3$ by low-mass halos with $M_h \!<\! 10^{10}$~{\Msun} ($L\!<\!{\onepLstar}$ according to Figure~\ref{fig:hmlr}) ejecting metals out to ${\Rg} \!\gtrsim\! 100$~kpc. Using cosmological hydrodynamic simulations, \citet{Wiersma10} reported that at least half of the mass in metals in the IGM at $z\!=\!2$ was ejected by halos with $M_h \!<\! 10^{11}$~{\Msun} ($0.01 \!\leq\! {\LLstar} \!\leq\! 0.1$), with even lower masses potentially dominating enrichment at higher redshifts. \citet{Scanna05} used a simple outflow model to show that the smallest, most numerous, and earliest sources were the most efficient at enriching the IGM, as they are able to enrich a larger comoving volume (corresponding to smaller physical volume) at early times. Similarly, \citet{Porciani05} found that the observed high cross-correlation of Lyman Break Galaxies (LBG) and {\CIV} absorbers \citep{Adelberger03} can be explained by IGM enrichment from $M_h \!=\! 10^{8}$--$10^{10}$~{\Msun} dwarf galaxies at $6 \!\leq\! z \!\leq\! 12$. From a detailed treatment including star formation, \citet{Samui08} corroborated these general results and also suggested that $M_h \!>\! 10^{10}$~{\Msun} halos  enrich a significant volume of the IGM only after $z\!=\!3$.

As gleaned from Figure~\ref{fig:fR}, our models suggest that for $\beta \!=\! 0.5$, high-mass halos have more extended weak {\CIV} envelopes (${\fR} \!>\! 1)$ than low-mass halos, a trend that is accentuated as redshift increases.  This is seemingly at odds with the expectation of efficient outflows and extended IGM enrichment at higher redshifts from low-mass halos as opposed to high-mass halos. In the case of gas radius being independent of galaxy luminosity ($\beta \!=\! 0$), our models suggest low-mass halos have more extended weak {\CIV} envelopes (${\fR} \!>\! 1)$ than high-mass halos.  However, this trend is accentuated as redshift decreases, which is again at odds with early IGM enrichment from low-mass halos.  

One possibility for the mismatch between observations, theory, and our models is that $\beta$ could evolve with redshift. As \citet{Chen01} has reported $\beta \!=\! 0.5$ for $z\!\leq\!1$, we could adopt that value at low redshift and this yields higher mass galaxies having larger gas radii at low redshift in our models. We can infer then, that if the luminosity scaling evolved such that $\beta\!=\!0$ at $ z\!\simeq\! 5$, our model predictions of ${\fR}({\LLstar},z)$ could, in principle, be aligned with current observations and theoretical expectations. However, there is no physical basis by which to assume $\beta$ evolves in this contrived fashion.

As stated in Section~\ref{sec:modelcontext}, our formalism treats each halo as isolated, and therefore does not allow for {\it overlap} between halos. \citet{Scanna05} showed how the clustering of intergalactic metals implies a volume overlap of individual enriched ``bubbles'' from galaxies. They posited that volume overlap may explain the large observed sizes of enriched regions, which would otherwise require extremely large energies to be ejected by a single source. It is well known that the clustering of galaxies relative to dark matter (the bias) increases with halo mass; high-mass halos are substantially more clustered than low-mass halos \citep[e.g.,][]{WT18}.  Thus, the volume of gas within or surrounding a massive halo may have been enriched by a galaxy other than the central galaxy in that halo. Significant overlap of gas envelopes from low-mass halos clustered around a high-mass halo would result in the smaller gas envelopes being ``absorbed'' into the gas envelope of the higher mass halo. Our formalism would then break down for high-mass halos, especially for higher redshifts where the physical volume of the universe was smaller and halo volume overlap was probably more severe.

There is a growing body of evidence suggesting that some phases of the CGM are more extended for galaxies in groups.  Galaxies residing in groups have a higher covering fraction of {\MgII} at $z\!\sim\!1$ \citep{Dutta20} and of {\CIV} at $z\!\sim\!3$ (Galbiati et~al., in prep) than isolated galaxies, implying higher volume filling factors in the CGM of group galaxies. Additionally, \citet{MF21} found that the majority of the {\CIV}+{\MgII} systems they observed at ${z\!=\!1\!-\!1.5}$ may be associated with multiple galaxies within 250 kpc of the quasar line of sight, suggesting absorbers might also exist in group environments. Similarly, \citet{Haman20} found most of their detected {\MgII} systems to be associated with pairs or groups of galaxies. \citet{nikki18} found that, compared to isolated galaxies, {\MgII} absorbers in group environments at $0.1 \!\leq\! z \!\leq\! 0.9$ have both a higher median {\EWr} and a higher covering fraction. They suggested that {\MgII}-bearing gas forms a kinematically complex intragroup medium in group environments.  These measurements of {\MgII} covering fractions led \citet{Lan20} to conclude that the evolution of cool gas may reflect the star formation activity of an entire halo consisting of the central galaxy and its satellites, as opposed to just the central.  Such ideas hark back to the original suggestions of \citet{York86}.

\citet{Chen10}, \citet{Bordoloi11}, and \citet{nikki18} all reported that the {\EWr} of {\MgII} absorbers in group environments does not decline with impact parameter out to about 200 kpc, unlike the very strong anti-correlation between {\EWr} and impact parameter observed for isolated galaxies. This fact is highly suggestive of gas halo overlap or an intragroup medium in overdense environments. \citet{lopez08} found an excess of strong {\MgII} absorbers and a paucity of weak {\MgII} absorbers (${\EWr} \leq 0.3~${\AA}) in $0.3\!\leq z\leq\! 0.9$ cluster environments, which they suggest is due to a population of absorbers associated with overdense cluster environments in which weak systems are destroyed (though they could be blended together).  

In both the FIRE-1 \citep{AA17} and FIRE-2 \citep{Hafen20} simulations, winds from satellite galaxies constitute an increasing fraction to the central galaxy CGM mass and this effect is more prominent with increasing halo mass of the central galaxy. In fact, \citet{Hafen20} reported that in $\sim\! 10^{12}$~{\Msun} halo progenitors, the central halo can actually accrete more gas from satellite winds than from recycled winds originating in the central galaxy at both $z\!=\!0.25$ and $z\!=\!2$.

However, for {\OVI} absorbers, \citet{Pointon17} found that the average {\EWr} in $z \!\leq\! 0.25 $ low-mass galaxy groups is smaller and the absorption velocity profiles narrower than for isolated galaxies.  They attributed this to higher virial temperatures in group environments (with more massive halos) that are able to ionize oxygen to higher-ionization states, thus diminishing {\OVI} absorption properties in groups. In general, {\OVI} absorption does not behave similarly to {\MgII} absorption in isolated galaxies, kinematically \citep{Nikki17,glenn19} or with halo mass \citep{Ng19}.

Crucially, the association of metals within the virial radius of a galaxy living a high-mass DM halo does not require that the metals originated in that galaxy. 
As \citet{Oppenheimer12} demonstrated, at high redshifts, metals ejected in winds in regions of lower overdensities will gravitate into denser regions of the IGM and progressively into higher mass DM halos with cosmic time. This ``outside-in'' enrichment of the CGM of higher mass halos results in higher ionization metals like {\CIV} tracing lower density, older gas. This gas would be unique to the ``ancient outflow'' gas seen in the simulation of \citet{Ford14}. It would be interesting to examine the simulations to determine whether this gas that has fallen in from the IGM is consistent with {\wvweak} {\CIV} absorbers.

The upshot is, highly extended envelopes of enriched gas surrounding galaxies should not reflexively be interpreted as evidence for powerful winds originating from the central galaxies. Material traveling at a constant speed of $500$~{\kms} for $300$ Myr would only be able to traverse up to $225$ comoving kpc at $z\!=\!2$, compared to $450$ comoving kpc at $z\!=\!5$.
Whereas these winds escape more easily from the potential wells of early low-mass halos, in higher mass halos the winds slow down due to hydrodynamic interactions with the denser CGM \citep[e.g.,][]{OD06}.
This is further theoretical corroboration that winds originating in higher mass halos may not reach the most extended regions of their own CGM. Of course, AGN outflows are believed to be an important mode of feedback in some massive galaxies \citep[e.g.,][]{Nelson19}, but the low space density of AGN, coupled with relatively brief duty cycles suggests that this mode of feedback is not nearly as prevalent as less energetic SN-driven winds (AGN do not dominate the measured {\dndx} of {\CIV} absorbers).

So instead of outflows emanating from $L\!\gtrsim\!{\Lstar}$ galaxies at various epochs, lower velocity outflows at $z\!>\!5$ transferred from the low-mass progenitors in the cosmic vicinity of such galaxies could explain the extended gaseous envelopes surrounding higher mass halos.
In fact, \citet{Diaz14} found evidence for this scenario; they demonstrated that their sample of {\CIV} absorbers at $z\!\sim\!5.7$ likely trace low-to-intermediate density environments of faint, low-mass galaxies, rather than massive ($L\!\geq\!{\Lstar}$) galaxies at that redshift.
This scenario would imply that the luminosity scaling parameter of $\beta\!=\!0.5$ in our models (which predicts highly extended gas radii for {\wvweak} {\CIV} absorbers at all redshifts, see Figure~\ref{fig:fR}), does not, in reality, describe a causal link between the central galaxy luminosity and the gas radius (relative to the virial radius). It would imply that, if the gas radius appears to behave in this fashion, the causality is due to enrichment from local lower mass halos clustered around the higher luminosity galaxy.

In view of these considerations, we are faced with an alternate explanation of  our $\beta\!=\!0.5$ prediction that high-mass halos have more extended {\CIV} envelopes than low-mass halos and that the most massive halos may have hosted {\wvweak} gas out to as much as twice their virial radius at $z\!\geq\!4$. If there is significant overlap between high-mass halos and their surrounding low-mass halos, it would render our model too simplistic and the definition of an isolated spherical ``gas envelope'' as meaningless. Halo overlap implies the importance of lower mass satellites and/or environmental effects on the CGM of massive galaxies and the nearby IGM.

\subsection{So, How Does {\CIV} Evolve?}
\label{sec:reconcile}

The preceding discussion leads to the emergence of a picture where low-mass dwarf galaxies at early epochs disperse carbon-enriched gas to the overdense regions surrounding massive central galaxies. This results in large, extended {\CIV} gas envelopes around massive, luminous galaxies, likely extending beyond their virial halos. 
The metals found at large distances (near the edge of, or outside, the virial radius) from their centers thus likely form in their satellites. 
This clearly poses a challenge for observational studies which attempt to derive constraints on galaxy outflows based on physical proximity of absorbers to a luminous galaxy, since absorbers detected at impact parameters may not have even originated inside the galaxy in question.

In this picture, gas in the outskirts of enriched gas halos are optically thin and mostly produced in ancient winds expelled from lower mass galaxies. Deeper in the halos, the gas becomes optically thick and/or kinematically complex and its origins lie in more recent outflows ejected by the more massive central galaxies. 
This picture is consistent with a steep scaling of gas radius with luminosity, even though the physical origin of gas at the edge of the halos may not be from the central galaxies themselves. Obtaining conclusive evidence of this scenario requires significantly increasing observations of gas radius across parameter space, supplementing the currently limited data described in Section~\ref{sec:compare}. One focus of this observational effort could be to target high-luminosity galaxies at high redshifts ($z\!>\!3$) so that the $\beta\!=\!0.5$ scenario can be tested at early epochs. Furthermore, simulations can be utilized to comprehensively characterize $\beta$ as a function of halo mass, redshift, {\wrlim}, and galaxy properties such as color, SFR, and environment. To the best of our knowledge, no such theoretical characterization of $\beta$ exists. An evolving or luminosity-dependent $\beta$ would muddy the waters of the physical picture we have outlined.

The evolution of {\CIV} gas halos tells a story of metal enrichment and changing ionization conditions across the vast majority of cosmic time. Optically thick {\CIV} envelopes approximately grow with the rise in metal enrichment deeper in the halo, caused by the rise in star formation activity across Cosmic Noon, and grow more gradually as metal enrichment slows down. In contrast, optically thin {\CIV} envelopes {\it decline} relative to virial halos across Cosmic Noon. Large extended envelopes are in place at high redshifts, but some of this gas transitions into virial halos, which grow steadily with cosmic time. The extragalactic UV background (UVB), which becomes harder with time from $z\!\sim\!5$ to a peak around $z\!\sim\!1$ \citep{HM12}, causes a change in carbon ionization fractions, and favors higher ionization states like {\CV} and {\CVI} as Cosmic Noon progresses \citep[][see their Figures 11 and 12]{Simcoe11}.

It is therefore reasonable that in the outskirts of gas halos, optically thin carbon is ionized to higher ionization states than {\CIV} between $z\!\sim\!5$ and $z\!\sim\!1$, thus reducing the relative sizes of the weak {\CIV} halos. 
Optically thick and/or kinematically complex {\CIV} in denser regions closer to galactic centers would be shielded from these ionization effects so that the strong absorbing halo sizes do not respond to these universal changes in background ionization conditions. 
The relative growth of {\wvweak} halos at $z\!<\!1$ is puzzling, nevertheless, as the gradually softening UVB favors lower ionization states of carbon than {\CIV} at these late times. 
We speculate that this growth may be due to increased collisional ionization of lower ionization carbon to {\CIV} in the outskirts of hot halos of massive galaxies at lower redshifts. Unfortunately, there is no strong empirical evidence for this idea currently.



\section{Conclusion}
\label{sec:conclude}

We investigated the statistical extent of metal-enriched gas relative to DM halos and their central galaxies by leveraging the observed statistics of {\CIV} absorption lines, DM halos, and luminous galaxies across $\sim$12.5~Gyr of cosmic time. 
We draw from the results of our \citetalias{H20} in which we developed a census of {\CIV} absorber populations with minimum limiting equivalent widths of ${\wrlim}\!=\!0.05$, $0.3$, and $0.6$~{\AA} over the redshift range $0 \!\leq\! z \!\leq\! 5$.
Here, we applied abundance matching using the observed UV luminosity functions \citep{Parsa16} and DM halo mass functions \citep{Murray13} to obtain a census of DM halos hosting $L\!\geq\!{\onepLstar}$ galaxies over the redshift range $0 \!\leq\! z \!\leq\! 5$.

\subsection{General Findings from Cumulative Statistics}

We first compared the cosmic incidence, i.e., the product of number density and cross-sectional area, of {\CIV} absorber populations with that of galaxy populations (see Figures~\ref{fig:dndx_halo+abs} and \ref{fig:fX}).

\begin{itemize}
\itemsep0em
\item $L\!\geq\!{\Lstar}$ galaxy halos alone cannot account for all the observed {\wvweak} {\CIV} absorbers, but they can account for all {\wweak} and {\wstrong} absorbers.

\item Assuming {\CIV} absorbers are co-spatial with DM halos, all absorbers live inside the virial radius of ${L\!\geq\!{\onepLstar}}$ and ${L\!\geq\!{\tenpLstar}}$ halos. Under this assumption, the weakest population of absorbers with {\wvweak} on average reside beyond the virial radius of $L\!\geq\!{\Lstar}$ halos, and inside the inner $\approx$30\% of $L\!\geq\!{\onepLstar}$ halos and inner $\approx$50\% of $L\!\geq\!{\tenpLstar}$ halos. Strong {\CIV} with {\wweak} and {\wstrong} typically resides in the inner $\approx$10\% ($\approx$50\%) of the virial halo of $L\!\geq\!{\onepLstar}$ and $L\!\geq\!{\tenpLstar}$ ($L\!\geq\!{\Lstar}$) galaxies. 

\item Accounting for lower mass halos (lower luminosity galaxies) results in {\CIV} absorbing gas residing further inside the inner regions of a halo and/or becoming less abundant than halos.

\item The incidence of {\wvweak} {\CIV} gas halos relative to that of virial halos decreases during the Cosmic Noon epoch to a minimum at $z\!\sim\!1$, regardless of minimum luminosity of the cumulative galaxy population. Following $z\!\sim\!1$ (over the last $\sim$8 Gyr), this relative incidence increases. On the contrary, the relative incidence of {\wweak} and {\wstrong} {\CIV} halos evolve very mildly, with some indications that they grow with cosmic time for $L\!\geq\!{\onepLstar}$ galaxy populations.

\end{itemize}

Next, assuming that each population of {\wrlim} absorbers has the same cosmic number density as $L\!\geq\!{\onepLstar}$ galaxies, we estimated the {\CIV} absorbing gas radius, {\Rg}, and the ratio of this gas radius relative to halo virial radius, {\fR}, for three {\CIV} absorber populations, ${\wrlim}\!=\!0.05$, $0.3$, and $0.6$~{\AA}, over the redshift range $0 \!\leq\! z \!\leq\! 5$ (see Figures~\ref{fig:lstar} and \ref{fig:10lstar}). 

\begin{itemize}
\itemsep0em
\item In {\Lstar} galaxy halos, the {\wweak} and {\wstrong} absorbing gas halo sizes are ${\fR}\!\sim\! 30$--$70$\% of the virial radius. The {\wvweak} gas halos are larger than virial halos (${\fR} \!>\! 1$) at $z\!=\!0$ and $z\!>\!2.5$, implying that some of the weakest {\CIV} absorbers may live in the IGM surrounding massive galaxies. For {\tenpLstar} halos, the relative gas radii are smaller, with ${\fR}\!\sim\!20$--$40$\% for {\wweak} and {\wstrong} gas and ${\fR} \!\sim\! 60$--$100$\% for {\wvweak} gas.

\item We find that {\fR} for {\wstrong} absorbers increases up to the end of Cosmic Noon ($z\!\sim\!1.5$) and then declines afterwards. In stark contrast, the relative radii of {\wvweak} gas halos decrease from $z\!\sim\!5$ to $z\!\sim\!1$ and increase again at $z\!<\!1$. The evolution of gas radii are very similar between {\tenpLstar} and {\Lstar} halos. 

\end{itemize}

To fully constrain relative gas radius as a function of luminosity and redshift, we require better constraints on the gas covering fraction $f_c(z)$ and luminosity-scaling of gas radius $\beta$. Because we have adopted halo abundance matching, larger galaxy luminosity implies larger halo mass, though the exact proportionality evolves with redshift (see Appendix~\ref{app:abundance}). 

\begin{itemize}
\itemsep0em
\item If we assume a redshift-independent value of the only available observational constraint of $\beta\!=\!0.5$ \citep{Chen01}, then we find that the relative gas radius for all absorber populations generally increases with galaxy luminosity (halo mass). In this scenario, the {\wvweak} gas radius exceeds the virial radius of massive $L\!\geq\!{\Lstar}$ halos at virtually all redshifts for $f_c \!=\! 0.5$ (see Figure~\ref{fig:fR}). 

\item If instead $\beta\!=\!0$, where the physical gas radius is not proportional to the central galaxy luminosity, the cumulative statistics yield a gas radius that decreases with increasing galaxy luminosity (halo mass).  In this case, {\wvweak} {\CIV} extends past virial radius for the least massive halos with $L\!\leq\!0.05{\Lstar}$ for $f_c\!=\!0.5$ at all redshifts. In both these scenarios, the {\wstrong} gas halos are always contained in the inner half of the virial halo. 

\end{itemize}

In summary, a strong scaling of gas radius with galaxy luminosity requires highly extended, super-virial gas structures around high-luminosity galaxies before Cosmic Noon and in the present day, while a luminosity-independent gas radius requires highly extended structures around high-luminosity galaxies after Cosmic Noon.

\subsection{State of the Observations}

We examined which of the two scenarios outlined above ($\beta \!\simeq\! 0$ or $\beta \!\simeq\! 0.5$) better agrees with the observed spatial location of gas as a function of galaxy luminosity, redshift, and $f_c$ (see Figure~\ref{fig:fR_comp}). We compared our model predictions of the ratio {\fR} with observed data for {\wvweak} {\CIV} at $z\!\simeq\!0$ \citep{Bordoloi14,Burchett16}, $z\!\simeq\!1$ \citep{MF21}, and $z\!\simeq\!2.5$ \citep{Rudie19}. Unfortunately, the small sample sizes and limitations in survey search space make it impossible to conclusively constrain $\beta$ and $f_c$ with the currently available data.

\begin{itemize}
\itemsep0em
\item Observational leverage would be most useful in constraining the model at high redshifts; no constraints on {\fR} exist at $z\!>\!3$, and at the Cosmic Noon epoch of $2 \!\leq\! z \!\leq\! 3$, we require observations at ${\fR} \!>\! 0.5$ for $L\!\geq\!{\Lstar}$ galaxies and at ${\fR} \!>\! 1$ for $L\!\leq\!{\tenpLstar}$ galaxies (both corresponding to impact parameters $D\!\gtrsim\!125$ pkpc). The latter observations are certainly more challenging to acquire, owing to the difficulty of locating faint dwarf galaxies at high redshift. 

\item At $1 \!\leq\! z \!\leq\! 1.5$, observations at ${\fR} \!>\! 1$ for $L\!\leq\!{\tenpLstar}$ galaxies ($D\!\gtrsim\!150$ pkpc) would be most useful, while at $z\!\simeq\!0$, observations at ${\fR} \!>\! 1$ for $L\!\sim\!{\Lstar}$ galaxies ($D\!\gtrsim\!400$ pkpc) and for $L\!\leq\!{\tenpLstar}$ galaxies ($D\!\gtrsim\!180$ pkpc) would be required. Theoretical simulations could also be of great utility in understanding the spatial distribution of {\wvweak} {\CIV} in these regions.

\end{itemize}

\subsection{Improving Our Understanding of {\CIV}}

From our analysis, we conclude that optically thin {\CIV} lives in the outskirts of virial halos (near or beyond {\Rv}) and in the IGM, while optically thick and/or kinematically complex {\CIV} lives in the inner half of virial halos.

\begin{itemize}
\itemsep0em

\item In the inner halos, {\CIV} gas is likely to be primarily produced via stellar feedback processes generating outflows which recycle back into the central galaxies. We suggest that the weak redshift evolution of optically thick and/or kinematically complex {\CIV} gas halos is mostly a consequence of the longer timescales than are seen for low-ionization gas, such as that traced by {\SiII} and {\MgII}. 

\item We propose that {\CIV} in the outer halos of galaxies is primarily produced in ancient outflows, likely originating in lower mass satellite galaxies at earlier epochs. 
The strong evolution in the relative gas radius ({\fR}) of {\wvweak} {\CIV} is very interesting and presents theoretical challenges.  This {\fR}  decreases from high redshift through Cosmic Noon and then grows rapidly from $z\!=\!1$ to $z\!=\!0$.  The cumulative statistics appear to favor a scenario in which {\wvweak} {\CIV} gas halos are more extended for more massive galaxies; this may in part be due to a population of lower mass satellites at earlier epochs that enrich their surroundings. 
\end{itemize}

We stress that the model we presented is but one of several possible scenarios of the connection between galaxies and their metal-enriched halo gas. To definitively discriminate between competing scenarios, we require a {\it significant\/} increase in observational data as well as improved characterization in theoretical simulations. Interestingly, our model and its implications for the baryon cycle could be also explored for the more thoroughly studied {\MgII} absorber-galaxy population, which has excellent measurements for {\dndx} to stringent {\EWr} sensitivity limits \citep[e.g.,][]{MS12, Zhu13, Vultures17, Chen17}.

Larger samples of observed galaxy-absorber pairs across a wide range of galaxy luminosities, impact parameters, and redshifts, will be crucial for constraining the nature of {\CIV} absorbers and their connection to galaxy evolution.  To the degree that our model applies, these observations would constrain $f_c(z)$ and $\beta$.

In the last few years, Multi-Object and Integral Field Spectroscopy-based programs like MEGAFLOW \citep[e.g.,][]{MF21}, KBSS \citep[e.g.,][]{Rudie19}, CGM at Cosmic Noon with KCWI \citep[e.g.,][]{nikki20}, MUDF \citep[e.g.,][]{MUDF19}, and MAGG \citep[e.g.,][]{Dutta20} have lead the way in mapping the CGM of galaxies across the vast majority of cosmic time and providing detailed information on the spatial distribution of gas halos relative to galaxies. In addition, measurements of galactic outflows at high redshifts, which will be enabled by the soon-to-be-launched {\it James Webb Space Telescope}, will be a direct empirical test of the scenario of early enrichment by low-mass halos we proposed. Finally, state-of-the-art cosmological hydrodynamical simulations such as EAGLE \citep[e.g.,][]{Schaye15}, FIRE-2 \citep[e.g.,][]{Hafen19}, and IllustrisTNG \citep[e.g.,][]{Nelson19} can be important tools in constraining the nature of highly uncertain model parameters such as $\beta$ across a broad range of physical properties and testing their sensitivity to adopted feedback prescriptions including outflow energetics, mass loading factors, and sources of feedback. These predictions can then be tested against observations as they become available.



\section*{Acknowledgements}


We would like thank Cameron Hummels and Peter Behroozi for valuable input and discussions that contributed to this work. We would also like to thank Benjamin Oppenheimer, Joop Schaye, Freeke van de Voort, and Dylan Nelson for insightful comments.
We are very grateful to Gwen Rudie, Ilane Schroetter, Nicolas Bouch{\'e}, and Joe Burchett for generously contributing unpublished quantities from their galaxy-absorber survey work for use in this paper. 
We also thank the anonymous referee for valuable comments which helped improve this manuscript.
N.M.N., G.G.K., and M.T.M.~acknowledge the support of the Australian Research Council through {\it Discovery Project} grant DP170103470. 
M.T.M. also thanks the Australian Research Council for Discovery Project grant DP130100568 for their partial support of this work.
Parts of this research were supported by the Australian Research Council Centre of Excellence for All Sky Astrophysics in 3 Dimensions (ASTRO 3D), through project number CE170100013.




\appendix

\section{Mapping Galaxies Onto Halos}
\label{app:abundance}

\begin{figure*}[htbp]
\centering
\includegraphics[width=0.8\textwidth]{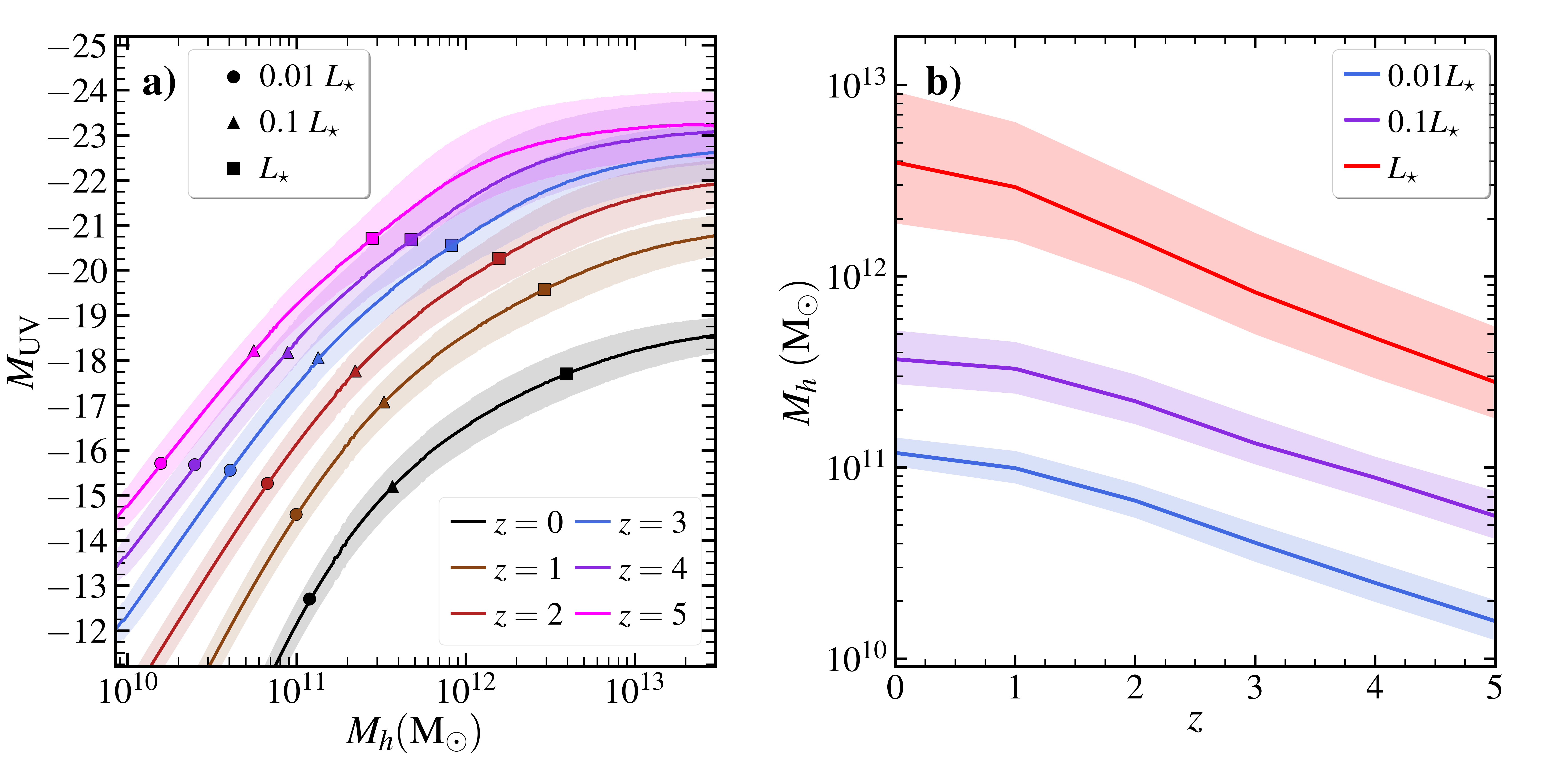} 
\vspace{-5pt}
\caption{Relationship between DM halo mass and galaxy luminosity, obtained from abundance matching. (a) The solid curves and shaded regions show the median UV magnitude-halo mass relations and $\pm1\sigma$ standard deviations, respectively, at a given redshift. On each curve, the circle, triangle, and square represents the magnitude and halo mass for a {\onepLstar}, {\tenpLstar}, and {\Lstar} galaxy, respectively. (b) Median halo mass corresponding to {\onepLstar} (blue), {\tenpLstar} (purple), and {\Lstar} (red) galaxies, as functions of redshift. The shaded regions represent the $\pm1\sigma$ standard deviations on the median.}
\label{fig:hmlr}
\end{figure*}

The DM halo mass function (HMF), $\phi(M_h,z)$, which describes the number density of DM halos per unit halo mass at a given redshift, can be integrated from a minimum halo mass {\Mhmin} to infinity, to obtain $n({\Mhmin},z)$, the cumulative number density of halos with $M_h \!\geq\! {\Mhmin}$. We used the code {\sc hmf}\footnote{\url{https://hmf.readthedocs.io/en/latest/index.html}} \citep{Murray13} to calculate the HMF at redshifts $0 \!\leq\! z \!\leq\! 5$, adopting the fitting functions of \citet{Behroozi13} for the halo multiplicity function. 
The choice of multiplicity function has a small effect ($\sim$1\%) on the abundances calculated.

Since galaxies populate halos, we expect there to be a mapping between halo and galaxy properties. While a number of observables have been used to probe the galaxy-halo connection, here we use galaxy luminosity $L$ to connect galaxies to their host halos. As stated in section~\ref{sec:galsdm}, this is primarily because historically, absorbing gas structures have been linked to the directly-observed luminosity of galaxies around which they are found \citep[e.g.,][]{Chen01,Chen10,Ribaudo11,nikki13b}. 

To that end, we apply the technique of abundance matching \citep[e.g.,][]{Kravtsov04,TG11,Behroozi13,Behroozi19,Klypin15}, where we match the cumulative number density of halos with ${M_h \!\geq\! {\Mhmin}}$ to that of galaxies with $L \!\geq\! {\Lmin}$,
\begin{equation} 
n(M_h\!\geq\!M_{h,\mathrm{min}},z) = n(L\!\geq\!L_{\mathrm{min}},z)\, .
\label{eq:abundance}
\end{equation}

This method allows us to compute the minimum halo mass corresponding to the minimum luminosity, thereby providing a link between galaxies and their host halos across cosmic time. Our operating assumption is that every DM halo more massive than {\Mhmin} hosts one galaxy more luminous than {\Lmin}. Similar to DM halos, the cumulative number density of galaxies with luminosities {\Lmin} and above, $n({\Lmin},z)$, is obtained by integrating the luminosity function (LF) $\phi(L,z)$ from {\Lmin} to infinity. We adopted the best-fit parameters of the 1500~{\AA} UV Schechter LF provided by \citet{Parsa16} for $z\!=\!0$--$8$. 
By abundance matching observed LFs to the HMFs, we generate halo mass-luminosity relations (HMLR) at redshifts $z\!=\!0$--$5$. Baryonic processes, such as the halo merger history, angular momentum relaxation, cooling, and stellar feedback can produce scatter in the HMLR. This scatter is often incorporated with a log-normal conditional luminosity function with a constant standard deviation $\sigma_M$ \citep[e.g.,][]{More09,Ren19,Whitler20}. 

To obtain the HMLR accounting for scatter in the abundance matching, we used the package {\sc AbundanceMatching}\footnote{\url{https://github.com/yymao/abundancematching}}, which utilizes an iterative deconvolution technique described by \citet{Behroozi10}. Following \citet{Whitler20}, we applied a standard deviation of $\sigma_M = 0.5$ in the HMLR (this value is approximately invariant with redshift). We ran $100,000$ random abundance matching realizations and extracted the median and standard deviation of UV magnitude, $M_{\mathrm{UV}}$, as functions of $M_h$, over redshifts $z\!=\!0$--$5$.  

The results of abundance matching are presented in Figure~\ref{fig:hmlr}. In Figure~\ref{fig:hmlr}(a), we present $M_{\mathrm{UV}}$ as a function of $M_h$ and $z$. The solid curves are the median HMLR and the shaded regions are the $\pm1\sigma$ standard deviation for a given redshift. On each curve, we indicate the halo mass and UV magnitude corresponding to {\onepLstar}, {\tenpLstar}, and {\Lstar} galaxies at that redshift. In Figure~\ref{fig:hmlr}(b), we show the redshift evolution of the median $M_h$ and its standard deviation for {\onepLstar}, {\tenpLstar}, and {\Lstar} galaxies. 

Due to the evolution in the HMF and LF, the HMLR also evolves. For instance, an {\Lstar} galaxy is only $\sim$0.5 magnitudes fainter at $z\!=\!2$ than at $\!z=\!5$, while a $\!z=\!0$ {\Lstar} galaxy is roughly $2.5$ mags fainter than its $z\!=\!1$ counterpart. The brightest galaxies show the steepest rise in halo mass. The halo mass for an {\Lstar} galaxy rises by $\sim$13 times from $z\!\sim\!5$ to $z\!\sim\!0$, while the halo mass for {\onepLstar} and {\tenpLstar} galaxies increase by $\sim\!7$--$8$ times. We also find that the HMLRs plateau at large halo masses ($M_h \!\gtrsim\! 10^{13}$~{\Msun}), and that the uncertainties in the HMLR increase with luminosity and redshift, as was noted previously \citep[e.g.,][]{Behroozi10,Whitler20}.

From abundance matching, we know the {\Mhmin} corresponding to a given {\Lmin} at any redshift. We can then compute the cosmic incidence, {\dndx}, of DM halos with $M_h \!\geq\! {\Mhmin}$ at redshift $z$, from Eq.~\ref{eq:dndx_abs} and Eq.~\ref{eq:nsighalo} for a fixed {\Lmin}. The virial radius $R_{v}(M_h,z)$ in Eq.~\ref{eq:nsighalo} is computed from the analytical approximation of \cite{Besla07},
\begin{equation}
R_{v}(M_h,z) =  \frac{206}{h^{2/3}} 
\left[ \left( \frac{M_h}{10^{12}} \right) 
\frac{97}{\Omega_{\hbox{\tiny M}}(z)~\Delta_{v}(z)} 
\right] ^{1/3} \, ,
\label{eq:rvir}
\end{equation}
where $R_{v}(M_h,z)$ is in kiloparsecs, $M_h$ is in solar masses, $\Omega_{\hbox{\tiny M}}(z)$ is the matter density of the universe at redshift $z$ and $\Delta_{v}(z)$ is the virial overdensity approximated using Eq.~6 of \citet{BN98}. 

\section{The Characteristic Gas Radius}
\label{app:rstar}

We estimate ${\Rstar}(z)$, the characteristic gas radius of an ${\Lstar}$ galaxy, by constraining the product $n(z)\sigma(z)$ for {\CIV} absorbers defined by $W_r \geq {\wrlim}$. Given that these absorbers have measured $dN/dX(z)$, we apply the relationship
\begin{equation}
\begin{array}{rcl}
\displaystyle
\frac{dN}{dX}(z) & \!\!=\!\!\! & \displaystyle \frac{c}{H_0} n(z) \sigma(z) \, , \\[15pt] 
\displaystyle
n(z) \sigma(z)  & \!\!=\!\!\! &  \displaystyle \bigintsss_{{\Lmin}/{\Lstar}}^{\infty} \!\!\!\!\!\!\!\!\!\!\!\!\!\!\!
\phi(\phi_{\star},t,\alpha,z) \, f_c(z) \pi R^2_{g}(t,z) \, dt 
\, ,
\label{eq:dndx_rgas}
\end{array}
\end{equation}
where $t={\LLstar}$, and where $\phi(\phi_{\star},t,\alpha,z)$ is the 1500~{\AA} UV luminosity function \citep{Parsa16}, {\Lmin} is the minimum galaxy luminosity, $f_c(z)$ is the covering fraction, and we assume gas absorbing cross-section is $\pi R^2_g({\LLstar},z)$, where $R_g({\LLstar},z) = {\Rstar}(z) ({\LLstar})^{\beta}$ (Eq.~\ref{eq:rg}). Here, we assume that the number density $n(z)$ of an absorbing population is equal to the number density of the population of galaxies with $L \!\geq\! {\Lmin}$. Integrating and solving for ${\Rstar}(z)$, we obtain
\begin{equation} 
R_{\star}(z) = 
\left[ 
\frac{H_0}{\pi c f_c(z)\phi_{\star}(z)}
\frac{dN/dX(z)}{\Gamma\left[x(z),  {\Lmin}/{\Lstar} \right]}
\right]^{1/2} \, ,
\label{eq:rstar}
\end{equation}
where $\Gamma[x(z),{\Lmin}/{\Lstar}]$ is the upper incomplete Gamma function \citep{Abramowitz72}, $x(z)\!=\! 2\beta\!+\!\alpha(z)\!+\!1$, and $\alpha(z)$ and $\phi_{\star}(z)$ are the faint-end slope and normalization, respectively, of the luminosity function at a given $z$.

\section{Observed gas radius data}
\label{app:obsdata}

For $z\!\simeq\!0$, we obtained the published impact parameter $D$, {\Rv}, and $r$-band luminosity {\LLstar} for COS-Dwarfs galaxies directly from \citet{Bordoloi14}, and $D$, {\Rv}, and halo mass $M_h$ for the \citet{Burchett16} sample from J.N. Burchett \citetext{2021, priv.\ comm.}. We applied a scale factor of $\sim\!1.2$ to both these {\Rv} to enable direct comparison with our chosen form of {\Rv}, since we do not use their overdensity definition of $\Delta_v = 200$ (see Appendix~\ref{app:abundance}). We also applied a scale factor of $\sim$1.2 to the halo masses of \citet{Burchett16} for the same reason. We converted these $M_h$ to UV {\LLstar} using the HMLR from abundance matching. Finally, we converted the \citet{Bordoloi14} $r$-band {\LLstar} to $1500$~{\AA} UV {\LLstar} by abundance matching the \citet{Parsa16} LF at $z\!=\!0$ to the $z\!=\!0$ SDSS $r$-band LF from \citet{Blanton03}.

For $z\!\sim\!1$, we obtained {\Rg} and stellar masses $M_{\star}$ of the galaxies in the \citet{MF21} sample from I. Schroetter and N. Bouch{\'e} \citetext{2021, priv.\ comm.}. The $M_{\star}$ were obtained from Spectral Energy Distribution (SED) fitting as described in \citet{MF19}. These masses were converted to $M_h$ using the stellar mass-halo mass relation (SHMR) parameterized by \citet{Girelli20}.  Specifically, we used the best-fit parameters of the $z\!=\!1$ SHMR with a relative scatter of 0.2 dex as presented in their Table 2. Their SHMR is broadly consistent with other published works \citep[e.g.,][]{Moster13,Behroozi13,Bocquet16}. Using these $M_h$ values, we then calculated {\Rv} from eq.~\ref{eq:rvir}, and used abundance matching to obtain {\LLstar}.  Two of the 11 galaxies had ${\LLstar} \!<\! 0.01 $ and were therefore below the minimum luminosity cutoff of our models.

For $z\!\sim\!2.5$, we obtained the published {\Rg} from \citet{Rudie19}, and unpublished apparent $\mathcal{R}$-band galaxy AB magnitudes from G. C. Rudie \citetext{2021, priv.\ comm.}. We converted the apparent magnitude to an absolute $R$-band magnitude with the average of $k$-corrections that assumes Sbc and Im galaxy SEDs \citep[see, e.g.][]{nikki13a}. These are reasonable assumptions for galaxy morphological type as the KBSS sample was composed of star-forming galaxies at $z\!\sim\!2.5$ \citep[e.g.,][]{Steidel14,Strom17}. We do not make extinction corrections as these are on the order of the uncertainty in the
$k$-correction \citep{Strom17}. Once again, we converted the $R$-band luminosity to the 1500~{\AA} UV band and obtained halo mass from abundance matching.




\bibliographystyle{aasjournal_nikki}
{\scriptsize \bibliography{Refs}}

\end{document}